\documentclass[useAMS,usenatbib]{mn2e}
\usepackage{graphicx}
\usepackage{latexsym}
\usepackage{psfig}
\usepackage{subfigure}
\title{The effect of stellar winds on the formation of a protocluster}

\author[J. E. Dale, I.A. Bonnell]{J. E. Dale$^{1,2}$\thanks{E-mail: jim@ig.cas.cz (JED)},
I. A. Bonnell$^{3}$\\
$^{1}$Astronomical Institute of the Academy of Sciences of the Czech Republic, Bocni II 1401/2a, 14131 Praha 4, Czech Republic\\
$^{2}$Department of Physics and Astronomy, University of Leicester, University Road, Leicester, LE1 7RH\\
$^{3}$Department of Physics and Astronomy, University of St Andrews, North Haugh, St Andrews, Fife, KY16 9SS}

\begin{document}

\pagerange{\pageref{firstpage}--\pageref{lastpage}} \pubyear{2006}

\maketitle

\label{firstpage}

\def\solmas{{M$_\odot$}}
\def\solm{{M_\odot}}
\def\mnras{MNRAS}
\def\apj{ApJ}
\def\aap{A\&A}
\def\apjl{ApJL}
\def\apjs{ApJS}
\def\bain{BAIN}
\def\pasp{PASP}
\def\araa{ARA\&A}
\def\ga{\sim}

\begin{abstract}
We present Smoothed Particle Hydrodynamics simulations of
protoclusters including the effects of the stellar winds from
massive stars. Using a
particle--injection method, we investigate the effect of
structure in close proximity to the wind sources and the short--timescale
influence of winds on protoclusters. We find that structures such as
disks and gaseous filaments have a strong collimating effect on winds. By a different technique of
injecting momentum from point sources into our simulations, we compare the large--scale and long--term effects of isotropic and intrinsically--collimated winds on protoclusters and find them to be similar, although the collimated winds take longer to exert a significant influence. We find that both types of wind are able to dramatically slow the global star formation process, but that the timescale on which they can expel significant quantities of mass from the cluster is rather long (approaching ten freefall times). Clusters may then experience rapid star formation very early in their lifetimes, before switching to a mode where gas is gradually expelled, while star formation proceeds much more slowly over many freefall times. This complicates any conclusions regarding slow star formation derived from measuring the star--formation efficiency per freefall time. We find that estimates of the efficacy of winds in dispersing clusters derived simply from the total wind momentum flux may not be very reliable. This is due to material being expelled from deep within stellar potential wells, ofen to velocities well in excess of the cluster escape velocity, and also to the loss of momentum flux through holes in the gas distribtuion. Winds have little effect on the accretion--driven stellar initial mass function (IMF) except at the very high--mass end, where they serve to prevent some of the most massive objects accreting more material. Feedback does not result in the formation of further massive stars through the monolithic collapse of massive cores. We also find that the morphology of the gas, the rapid motions of the wind sources and the action of large--scale accretion flows prevent the formation of bubble--like structures. These effects may make it difficult to discern the influence of winds on very young clusters.\\
\end{abstract}

 
\section{Introduction}
Elucidating the effects of stellar feedback on embedded star clusters is a problem of key importance in astrophysics. $\sim90\%$ of newborn stellar systems do not survive their embedded phase to become open clusters (\cite{2003ARA&A..41...57L}). In addition, star formation at the scale of molecular clouds and embedded clusters is not a very efficient process (e.g. \cite{1991IAUS..147..275F}). There are two possible reasons for these observations. If star--forming complexes are initially gravitationally bound (as is usually assumed), some process must expel most of the molecular gas before it can be incorporated into stars, while also usually disrupting the system. Alternatively, if such complexes are not initially bound, the low star formation efficiency and low survival probability are a natural consequence of this, although bound stellar systems can be left behind in some cases (\cite{2004MNRAS.347L..36C}, \cite{2005MNRAS.361....2C}).\\
\indent If star--forming systems are initially bound, an obvious candidate for a means of unbinding them is feedback from their own stars. In embedded clusters massive enough to form O--type stars (i.e. those with masses $>10^{3}$ M$_{\odot}$, the ionizing radiation, winds and eventual supernova (SN) explosions of these stars are likely to dominate the combined effect of the (much more numerous) lower mass objects. Photoionization and winds are particularly important in this respect, since they act for the duration of the O--star's lifetime, whereas SNe cannot occur for at least $5$ Myr after the onset of star formation. This corresponds to the freefall time of a molecular cloud of number density $\sim10^{2}$ cm$^{-3}$, approximately the minimum density required to form molecular gas able to cool efficiently and form stars (\cite{2001ApJ...562..852H}). If it is the case that star formation in, and disruption of, molecular clouds occurs on timescales of around the local freefall time (\cite{2000ApJ...530..277E}, but see \cite{2007ApJ...654..304K}), SNe may occur too late, particularly as the density of a molecular cloud generally increases considerably (with a corresponding decrease in the freefall time) before any massive stars are able to form.\\
\indent 
In this paper, we consider the effects of the winds from massive stars on the cluster in which they form. In reality, winds and photoionization act in concert and it is not clear which will dominate, nor if one will help or hinder the other (see \cite{2001PASP..113..677C} for a discussion of this subject). In this paper, we model winds alone -- we leave a comparative study including winds and ionization to a later date.\\
\indent 
The interaction of a stellar wind with the interstellar medium (ISM) has been modeled analytically and numerically by several authors, although usually in one--dimensional geometries. \cite{1975ApJ...198..575S} and \cite{1977ApJ...218..377W} introduced the idea of dividing the wind bubble into several concentric zones. The innermost zone in their models consists of freely--flowing wind and the outermost zone of undisturbed ISM. The number and contents of the zones in between is the subject of considerable debate. \cite{1975ApJ...198..575S} have a single zone between the free--flowing wind and the undisturbed ISM, consisting of a shocked shell driven by the momentum of the wind. It is not difficult to show that this assumption produces a wind bubble whose radius $R(t)$ increases with time as $R(t)\propto t^{\frac{1}{2}}$ \citep[e.g.][]{1988RvMP...60....1O,1999isw..book.....L}). \cite{1977ApJ...218..377W} introduce two zones in this region -- an inner shell of hot shocked wind and an outer shell of shocked ISM separated by a contact discontinuity. The difference between the models lies in Steigman et al's assumption that the shocked wind can cool and that the bubble is momentum--driven, as opposed to Weaver et al's contention that the shocked wind cannot cool and that the bubble is thus pressure--driven (resulting in a radial evolution of the form $R(t)\propto t^{\frac{3}{5}}$). \cite{2001PASP..113..677C} discuss at some length the validity of these assumptions, which essentially rests on whether the contact discontinuity between the shocked wind and the shocked ISM is stable, which governs how likely mixing and subsequent cooling of these phases is. Analytical work by \cite{1983ApJ...274..152V} and simulations by \cite{1996A&A...316..133G} suggest that the discontinuity may not be stable and the shocked wind may therefore be able to cool efficiently. In this case, the wind bubble is driven purely by ram pressure and $R(t)\propto t^{\frac{1}{2}}$.\\
\section{Numerical methods}
The code used in these calculations is a hybrid SPH/N-body code described in \cite{1995MNRAS.277..362B}. Hydrodynamic equations are solved using the SPH formalism and stars are modeled as collisionless point masses which are able to accrete gas particles.
We have included the effects of stellar winds emanating from point
masses in two ways. In the first instance, we model the wind directly
by injecting low--mass SPH particles. The second method models the
sink particles as momentum sources and distributes the momentum flux
amongst the SPH gas particles near the wind sources. 

\subsection{Particle winds}

\begin{figure*}
     \centering
     \subfigure[Shock radius, uniform medium ($\rho=$constant). Solid line is the analytic fit $R(t)\propto t^{1/2}$, circles are the results from the simulation.]{
          \label{fig:part_uniform}
                
          \includegraphics[width=.30\textwidth]{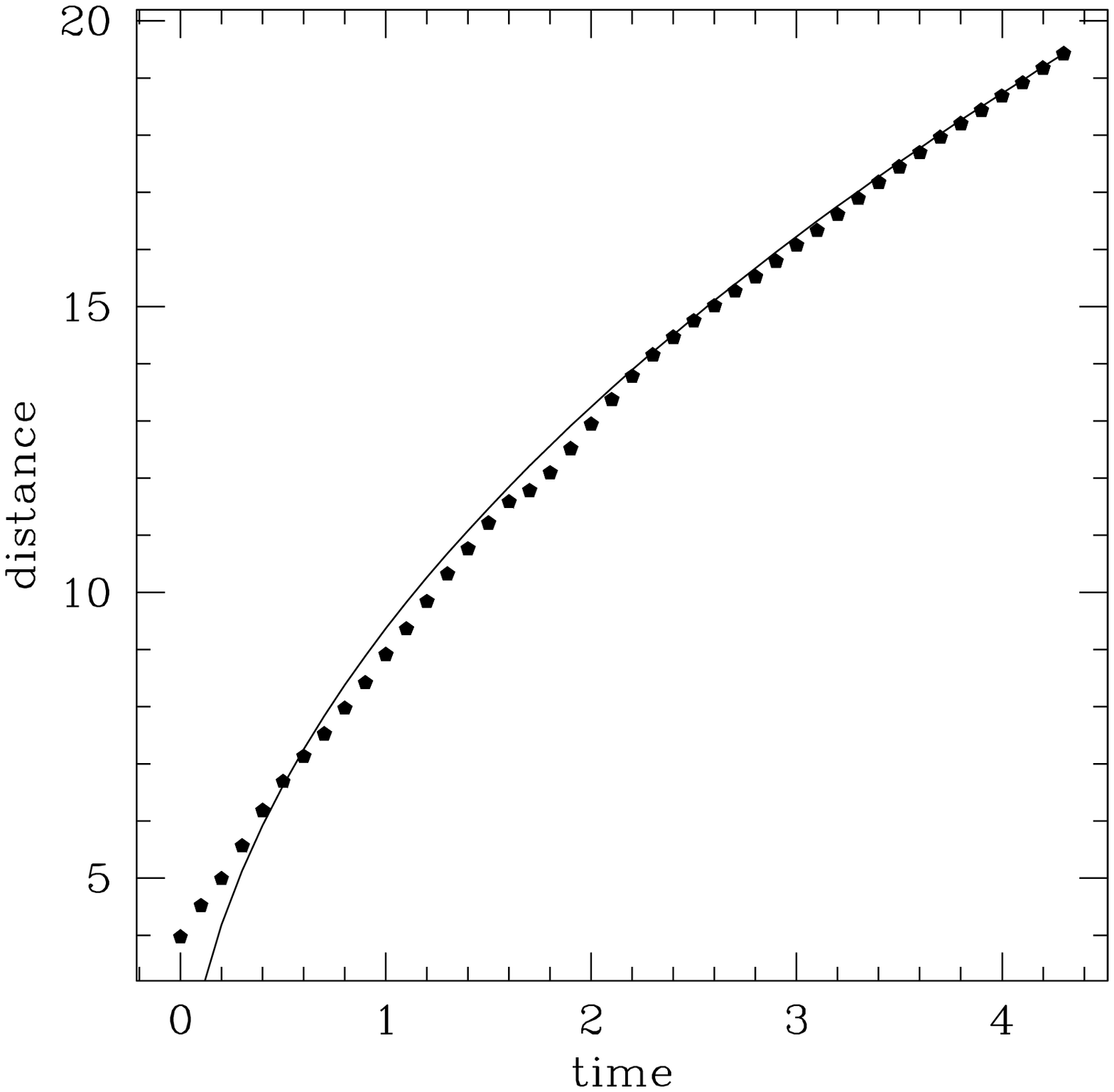}}
     \hspace{.1in}
     \subfigure[Shock $y$--displacement, slab with $\rho\propto y^{-1}$. Solid line is the analytic fit $Y(t)\propto t^{2/3}$, crosses are results from the simulation.]{
          \label{fig:part_slab}
          \includegraphics[width=.30\textwidth]{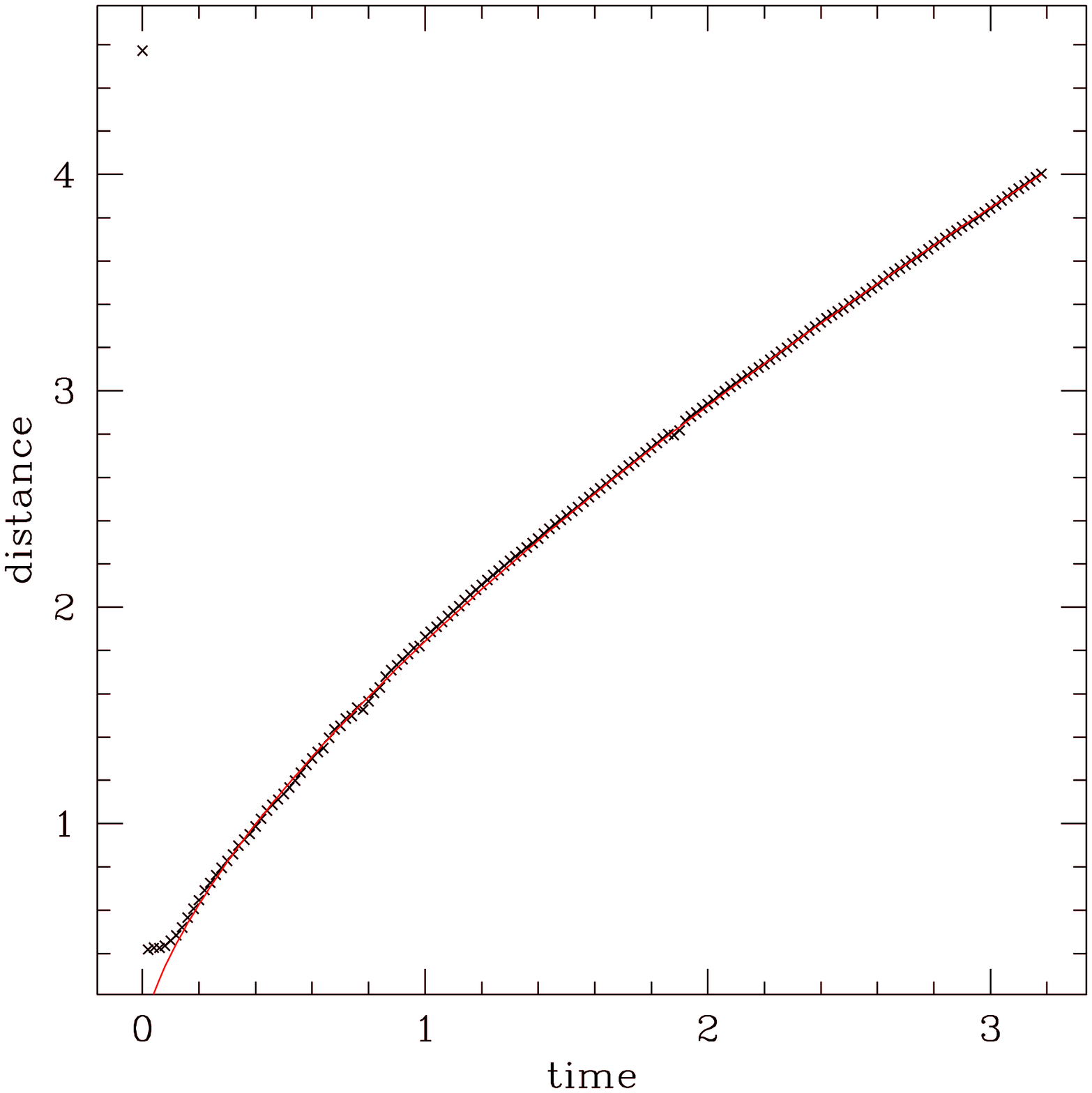}}
      \hspace{.1in}
      \subfigure[Particle plot of particles within $2h$ of the $x$--plane ($y$ on the vertical axis, $z$ on the horizontal axis), slab with $\rho\propto |y|^{-1}$. Blue crosses are regular gas particles, red particles are wind particles.]{
          \label{fig:part_slab_parts}
          \includegraphics[width=.30\textwidth]{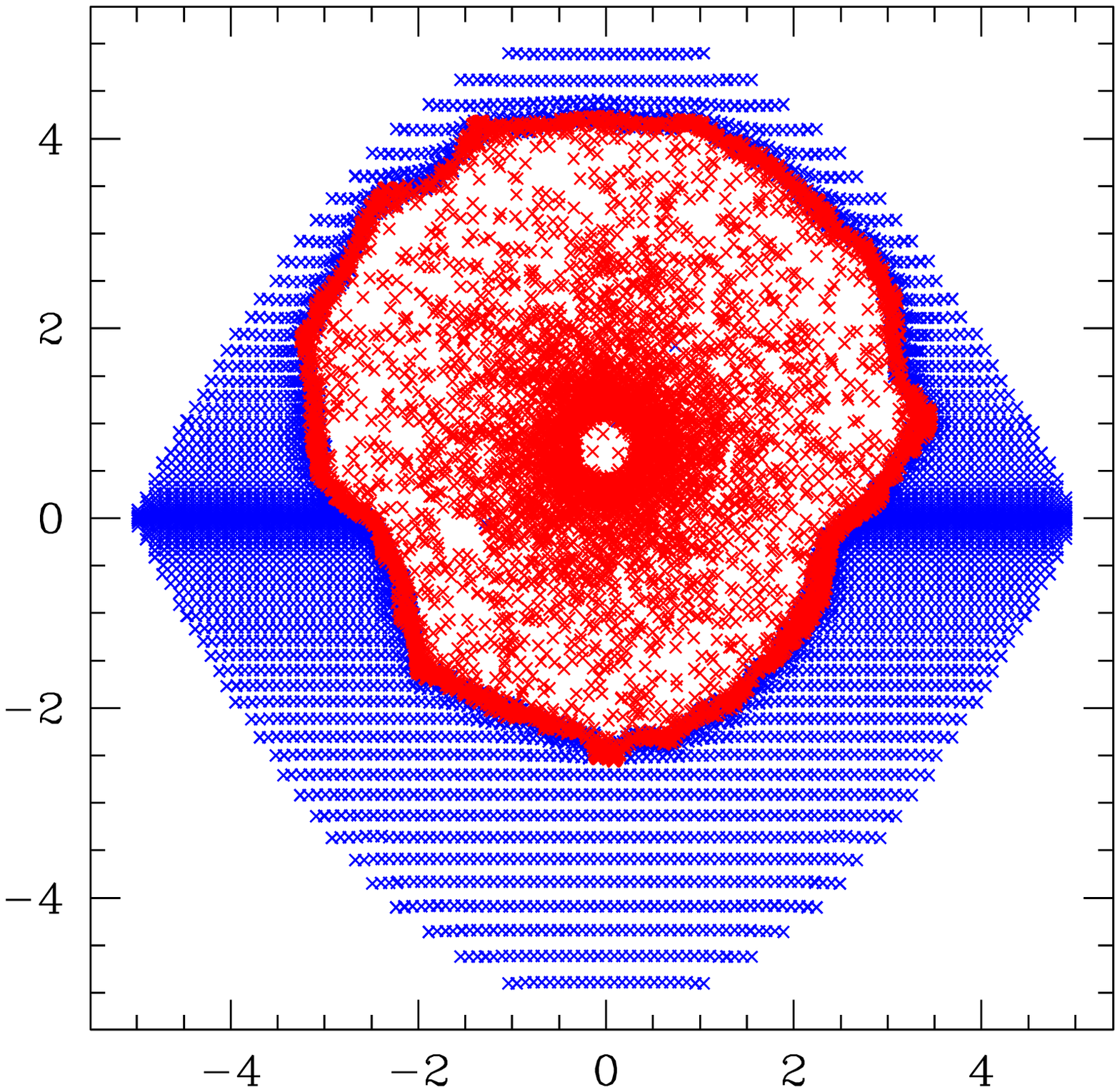}}

     \caption{Results of tests of the particle--injection wind code. Quantities are in code units.}
     \label{fig:part_tests}
\end{figure*}

\begin{figure}
\centerline{
\psfig{figure=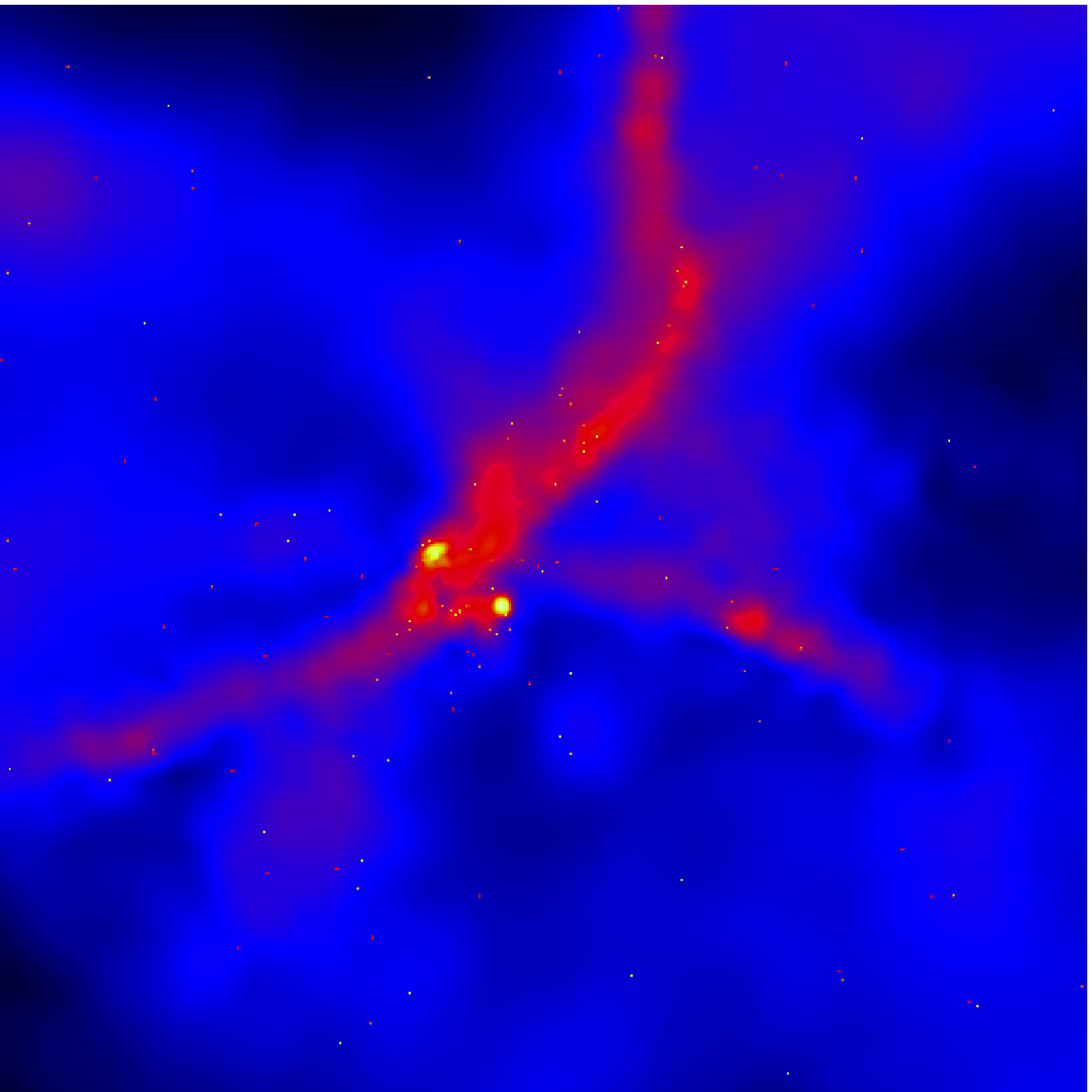,width=4.truecm,height=4.truecm}
\psfig{figure=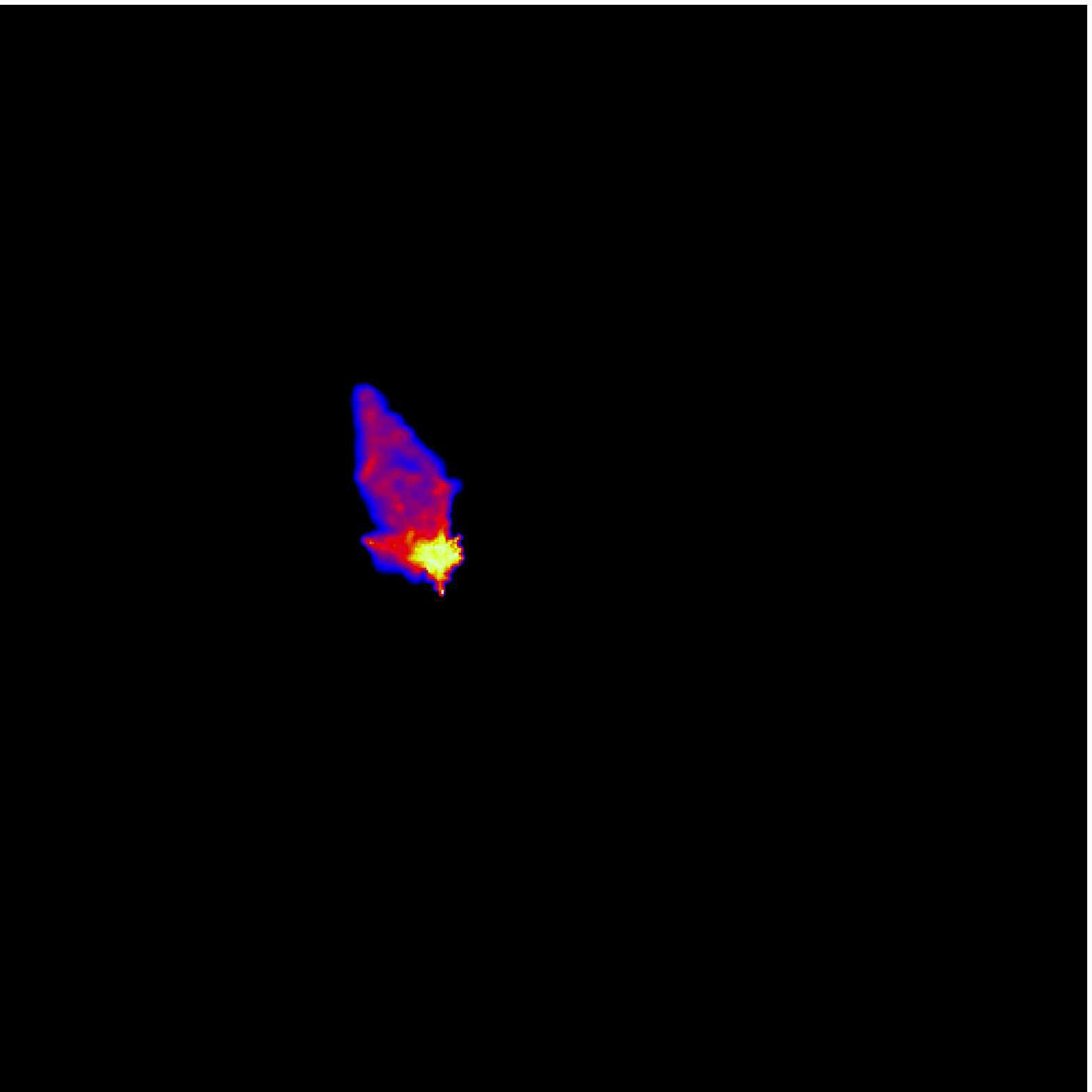,width=4.truecm,height=4.truecm}}
\centerline{
\psfig{figure=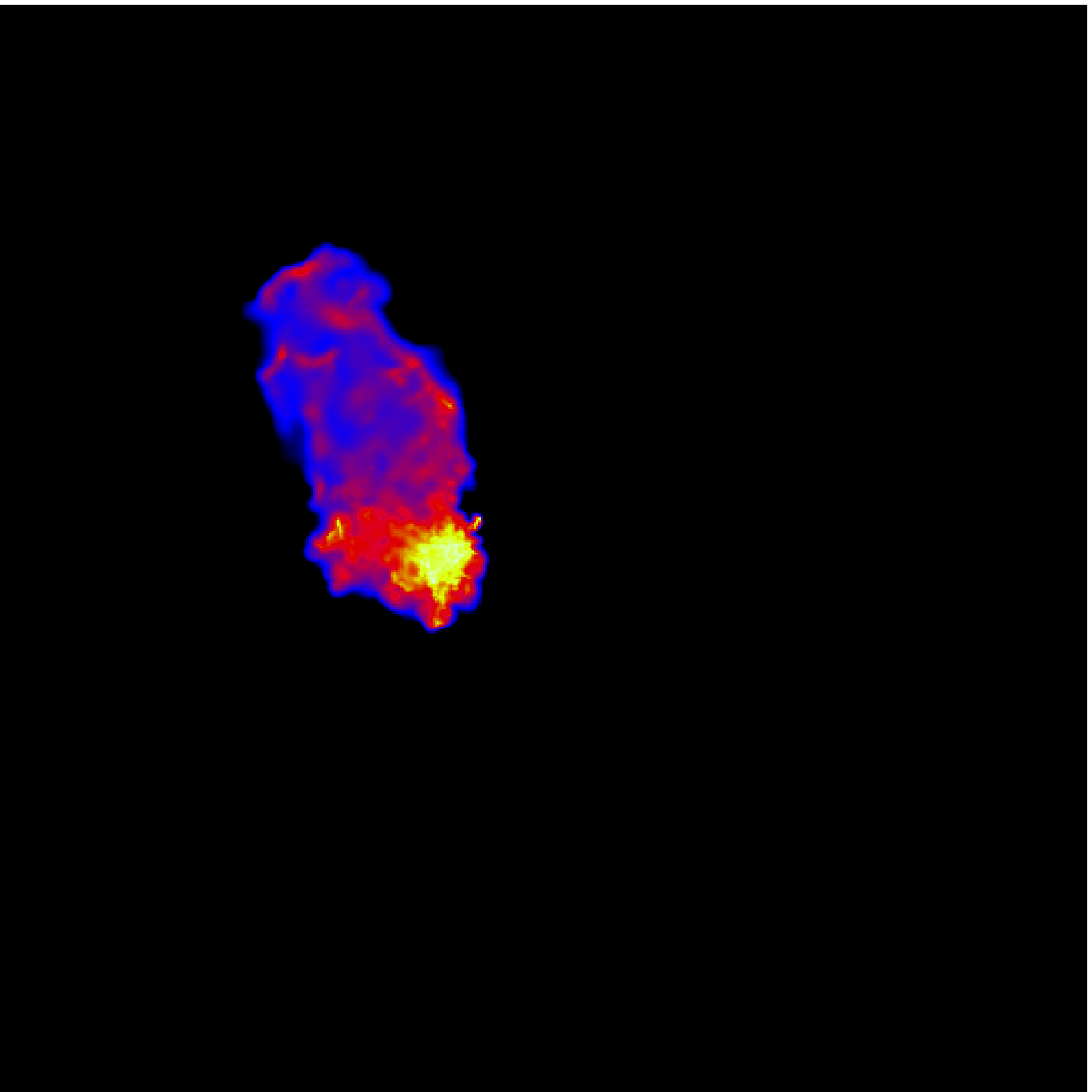,width=4.truecm,height=4.truecm}
\psfig{figure=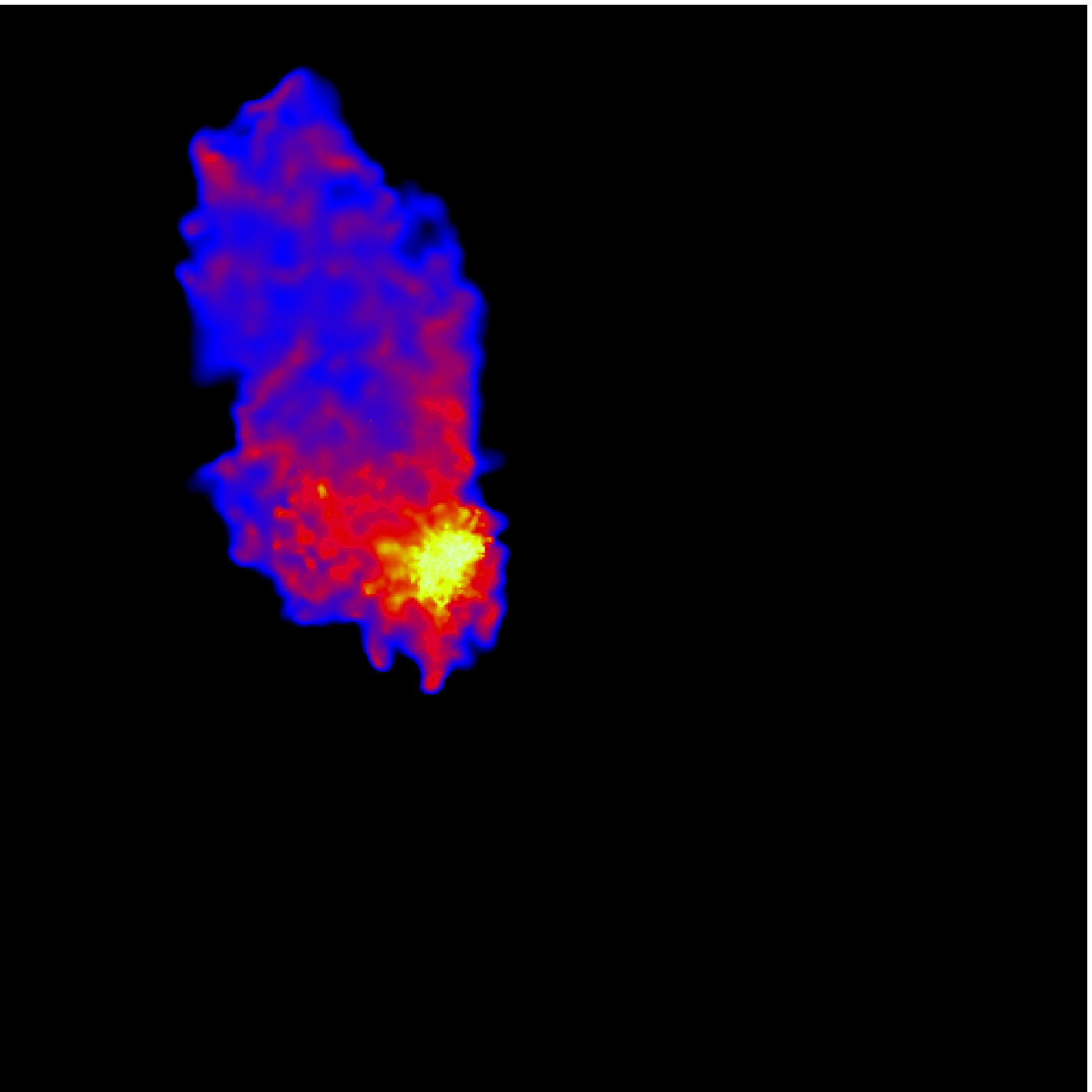,width=4.truecm,height=4.truecm}}
\caption{\label{outflowBB02} The outflowing gas from a massive star's wind is shown (panels 2-4) as it
is collimated by the filamentary structure in the surrounding cluster
gas (far left panel). Images are $0.05$ pc ($\approx10^{4}$ au) on a side. The column density of the gas is shown from 1.5 to 400 g cm$^{-2}$ in the top left panel
while the column density of the wind extends from $10^{-5}$ to $0.02$ g cm$^{-2}$.}
\end{figure}

The first method of handling stellar winds involves injecting low--mass SPH particles
from the chosen sink-particles.  
In order to establish a continuous wind from the ejection of discrete particles, a series
of shells are emitted at regular time intervals $\Delta t$. Each shell consists of $n_{\rm wind}$ 
(typically 300-500) SPH
particles which are chosen randomly from 6141 positions equally spaced over the surface of a sphere, with random rotations between sucessive shells to ensure that particles do not cluster along any particular directions. The shells radial width is determined from the wind velocity, $v_{\infty}$,
and the time interval between each shell's ejection:  $\Delta r=v_{\infty}\Delta t$.
The $r$--coordinates of the particles injected into the innermost shell are chosen randomly in the range $\left[r_{min},r_{min}+\Delta r\right]$, where $r_{min}$ must be outside the accretion radius of the sink particle. The mass of the wind particles are then set to provide the desired mass loss rate.
Provided that there are a reasonable number ($>10$) shells between the wind source and the nearest regular gas particle, this procedure produces an intrinsically smooth wind. However, owing to to the very high wind terminal velocity, this dictates that the timestep $\Delta t$ on which winds must be updated be very small, which makes the code comparatively slow.\\
\indent In order that the interaction of wind particles with regular gas particles at the contact surface is well modelled, the smoothing lengths of the wind particles should be similar to  
those of the regular gas particles, so that both types of particle should have $50-100$ neighbours of either particle type. Once a contact surface with an inner skin of wind particles is established, this is achieved reasonably well by the SPH code without assistance. However, when the wind is first launched, to avoid particle intepenetration at the contact surface, the smoothing lengths of the wind particles must be set so that they are similar to those of the regular gas particles with which they will first come into contact. We first define a smoothing length $h_{\rm near}$ and a distance $R_{\rm near}$ representative of the regular particles nearest the wind source from the \textit{initial} gas distibution. Since the smoothing lengths of the wind particles evolve in the spherical wind as $h \propto r^{2/3}$, the wind-particles' smoothing lengths can be set so that $h_{\rm wind}(R_{\rm near})\approx h_{\rm near})$ by an appropriate combination of are then set initially from a combination of  $n_{\rm wind}, \Delta r$ (determined by the choice of $\Delta t$) and $r_{\rm min}$. We set $r_{\rm min}$ to be just outside the accretion radius of the wind source ($10^{-4}$pc in all the simulations in this paper) and choose $n_{\rm wind}$ to be 300--500 to ensure that individual wind shells are reasonably smooth, and use this to constrain $\Delta t$. In the simulations presented here, these quantities are estimated at the beginning of each run and kept constant. It would improve the code to set their values dynamically, but we find in the simple test cases below and the more challenging simulations presented later that particle interpenetration is minimal and that the shock surfaces are well modelled. Once a shock has been well established, particle interpenetration is less of a problem, since the the dense particles at the contact surface prevent it. Another improvement to the method, not implemented here, would require that the smoothing lengths are calculated only considering particles of the same type
(i.e. wind particles must have $\sim50$ wind particle neighbours, and regular particles must have $\sim50$ regular neighbours). When particles of very different properties (masses, temperatures etc) are present in the same simulation, the evolution of the smoothing lengths should consider
the number of particles with similar properties. In the simulations involving highly inhomogenous gas distributions, this would help
limit potential problems due to lower density regions being under-resolved and thus
the artificial creation of voids by the winds.\\
\indent The particle-injection wind code was tested in two simple cases: (i) a single wind source in a uniform medium (ii) a single source embedded in a slab whose density profile follows $\rho\propto |y|^{-1}$. The analytic solution for the radius of the wind bubble in the former case, $R(t)\propto t^{1/2}$, can be found in, e.g, \cite{1988RvMP...60....1O}. In the latter case, we consider a disc perpendicular to the $y$--axis and of constant size plouging through the density distribution and sweeping up mass proportional to $\int_{0}^{Y}\rho(y)$d$y$. The flux absorbed by the disc is proportiona to $y^{-2}$, so the size of the wind bubble in the $y$--direction as $Y(t)\propto t^{2/3}$. In Figures \ref{fig:part_uniform} and \ref{fig:part_slab}, we show the results of these tests compared to the appropriate analytic functions. Figure \ref{fig:part_slab_parts} shows a particle plot from the simulation of the wind source embedded in a slab, and demonstrates that there is no interparticle penetration. We conclude that the particle--injection code performs well in these simple test cases.\\
\indent In Figure~\ref{outflowBB02}, we show the results of the evolution of a wind bubble in a highly inhomogenous medium. We use the standard SPH viscosity but with
the values $\alpha=2$ and $\beta=4$ in order to ensure the shock
is adequately captured. The wind particles are hot ($10^{4}$K, corresponding to an ionised wind) and are assumed to behave isothermally. We again find that there is no interparticle penetration.
The great difficulty in using SPH particles to model the stellar winds is that their
dynamical timescales are much shorter than that of the other gas particles present in the simulation.
This means that the evolution can be followed only for relatively short timescales and long term
effects cannot be investigated. This is the principal reason why the
second method, described below, has also been developed.\\
\subsection{Momentum--injection winds}
\begin{figure}
\includegraphics[width=0.47\textwidth]{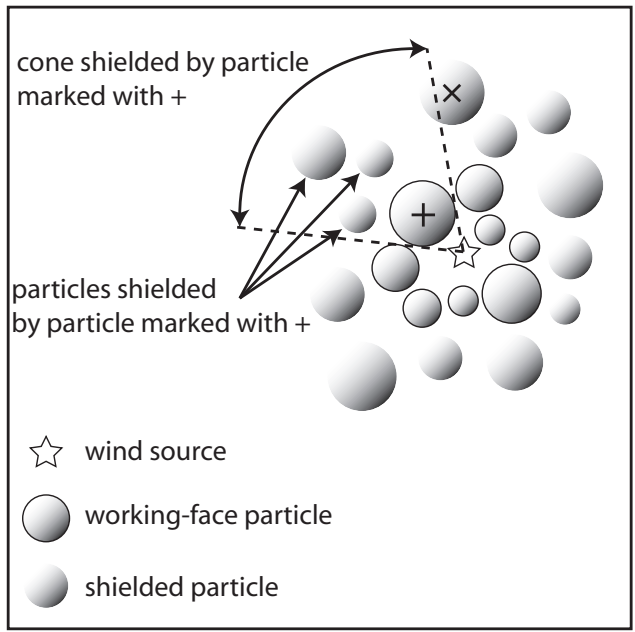}
\caption{Illustration of the working face. Note that the particle marked with a $\times$ symbol is not shielded by the working face particle marked with a + as that working face particle does not cut the line joining the wind source to the centre of the particle marked with $\times$.}
\label{fig:face}
\end{figure}

\begin{figure*}
     \centering
     \subfigure[Shock radius, uniform medium ($\rho=$constant). Dashed line is the analytic fit $R(t)\propto t^{1/2}$, solid line is the result from the simulation.]{
          \label{fig:mom_uniform}
                
          \includegraphics[width=.30\textwidth]{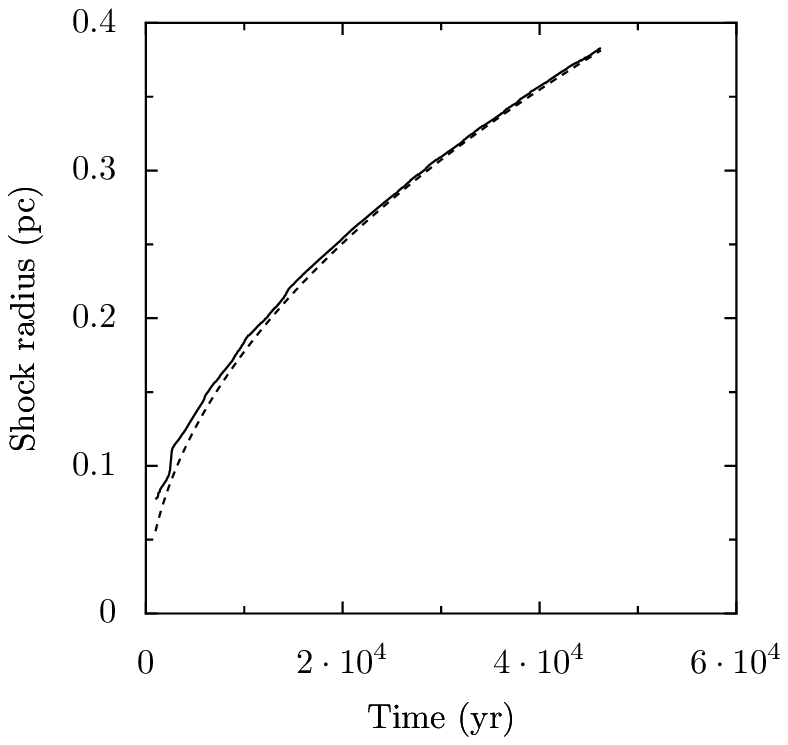}}
     \hspace{.1in}
     \subfigure[Shock radius, medium with $\rho\propto r^{-1}$. Dashed line is the analytic fit $R(t)\propto t^{2/3}$, solid line is the result from the simulation.]{
          \label{fig:mom_oneover}
          \includegraphics[width=.30\textwidth]{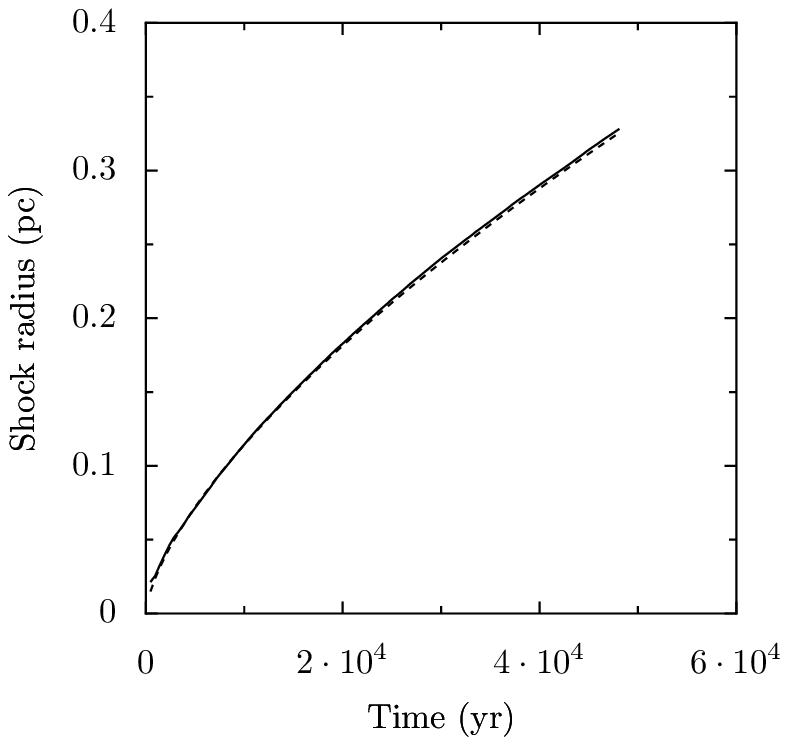}}
      \hspace{.1in}
      \subfigure[Shock $y$--displacement, slab with $\rho\propto |y|^{-1/2}$. Dashed line is the analytic fit $Y(t)\propto t^{4/7}$, solid line is the result from the simulation.]{
          \label{fig:mom_slab}
          \includegraphics[width=.30\textwidth]{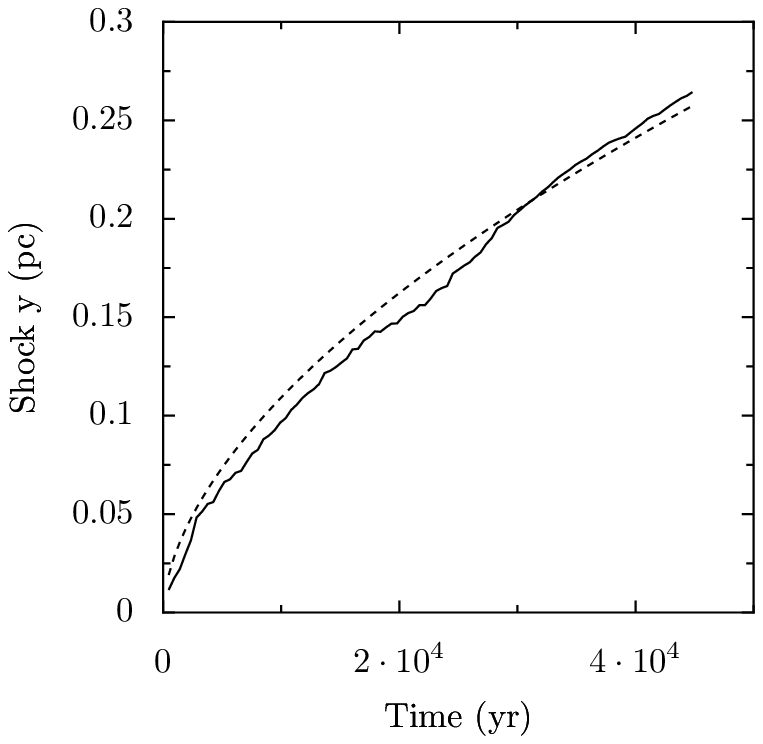}}

     \caption{Results of tests of the momentum--injection wind code.}
     \label{fig:mom_tests}
\end{figure*}

The second method is to inject momentum into the simulation and
distribute it amongst the regular SPH gas particles. This has the
advantage that fast--moving wind particles, with corresponding small
timesteps, need not be introduced into the simulation.
This allowed us to follow the evolution of the wind bubbes for much longer times, so that we could study the effects of winds on the early evolution of protoclusters. \\
\indent In line with the work of \cite{1983ApJ...274..152V} and
\cite{1996A&A...316..133G}, we consider winds driven purely by ram
pressure (although it must be admitted that this choice is also partly
motivated by computational necessity, since simulation of
pressure--driven winds requires detailed radiative transfer
calculations which are beyond the scope of this work).\\
\indent Each source is first assigned a momentum flux. The SPH
particles around the source over which this flux is to be distributed are then identified. We define around each wind source a `working face' (by analogy with a coal face) consisting of the minimum number of SPH particles which are required to shadow or shield \textit{all} the particles in the simulation from the wind source. In order to shield a given particle, a working--face particle must cut a line joining that particle's centre to the wind source. A momentum packet leaving the source must then either strike a working--face particle (before striking any other particles) or strike no particles at all and exit the cloud In which case the working face does not cover all of $4\pi$ steradians). We illustrate this idea in Figure \ref{fig:face}.\\
\indent Once the working face is defined, the momentum flux received by each working face particle must be determined. Since most working face particles are partially obscured by others, the fraction of the source's momentum flux absorbed by each one is not simply dictated by the solid angle the particle subtends at the source. What is required instead is the \textit{unobscured} solid angle subtended by each working face particle. In principle, this could be calculated geometrically by considering the particles as overlapping circles (seen from the point of view of the souce). However, this problem becomes very complicated when there are many such mutually overlapping circles of different sizes, as will be the case in a complex density distribution with SPH particles of different sizes at different distances from the source. We therefore do not explicitly calculate the unobscured solid angle presented to the source by the working face particles. Instead, we estimate the flux they each receive using a method analagous to that employed by Monte Carlo radiative transfer codes, although the problem at hand is much simpler. We imagine that the wind momentum flux is divided up into a number of packets which are emitted in random directions from the wind source. Each packet either strikes an SPH particle or does not -- it is likely in the highly--inhomogeneous gas distributions existing in molecular clouds that some stellar wind exits through a hole in the cloud and strikes the interstellar medium outside, but we do not concern ourselves with this process. We assume that if a momentum packet strikes an SPH particle, all of the momentum in the packet is absorbed by that particle. If a packet does not strike any particles, we ignore it (although the momentum of the packet is still debited from the source's total momentum flux).\\
\indent Up to ten passes are performed in which $100\times n_{face}$ packets are launched, where $n_{face}$ is the number of particles in the working face. The wind code exits when either all particles in the working face have been hit at least once, or ten passes are complete. Note that, owing to partial shadowing by other working face particles, the proportion of hits any member of the working face is likely to receive is highly variable, with a few particles very close to the wind source likely to be struck many times, while those far from the source which are partially shielded by other particles may only receive a few impacts. It is this issue that necissitates the use of the Monte Carlo method, since the momentum flux recieved by a working face particle is not simply. The momentum flux received by each particle in the working face is then determined by multiplying the wind flux from the source by the fraction of emitted packets hitting that particle.\\
\indent We considered three test cases: (i) a single wind source in a uniform medium (ii) a single source in a spherically symmetric density distribution following $\rho\propto r^{-1}$ (iii) a single source embedded in a slab whose density profile follows $\rho\propto |y|^{1/2}$. Referring again to \cite{1988RvMP...60....1O}, we find that the wind bubbles in the first two cases follow $R(t)\propto t^{1/2}$ and $R(t)\propto t^{2/3}$. In the third case, we again consider a disc of constant size moving in the $y$--direction and obtain the size of the wind bubble in the $y$--direction as $Y(t)\propto t^{4/7}$. We show the results of these tests in Figure \ref{fig:mom_tests} and compare them to appropriate analytic solutions. In each case, the extent of the wind bubble was followed by tracking the positions of the densest gas particles, i.e. those in the shock driven by the wind. The slight jaggedness in the plot of the simulated data in Figure \ref{fig:mom_slab} is a result of the necessity of using only a few tens of particles to accurately track the tips of the wind bubble in the positive and negative $y$--directions. We find that the code reproduces the expected behaviour very well.\\ 
\section{Intrinsic and extrinsic collimation}
Given the complex structures of molecular clouds, it is likely that an
isotropic wind emitted by a star will be collimated to some degree by
density inhomogeneities near the star. This effect has been observed
in simulations of the emission of photoionizing radiation from
O--stars by \cite{2005MNRAS.358..291D}, and we observe it in our
particle--injection wind simulations, as detailed in Section 4, where
we show that small--scale structure very close to the wind sources can collimate outflows into bipolar or even unipolar
morphologies. For example, in Figure~\ref{windMS1} the presence of a
circumstellar disc results in  a wind bubble formed that resembles an
hourglass, rather than a sphere. We refer to such small--scale
collimation as 'intrinsic' collimation.\\
\indent In our long--timescale protocluster simulations, in which we
use the momentum--injection technique to model winds, it is likely
that the winds would be intrinsically collimated by small--scale
structure close to the wind sources which is not well resolved. To
investigate the effect this might have on the evolution of the
protocluster, we include the effects of intrinsic collimation by modifying the angular dependence of the momentum output. We consider primarily collimation due to a circumstellar disc.  Although we may not be able to resolve the disk itself, our code does calculate the angular momentum of sink particles, derived from the angular momentum of the material from which the sink originally formed, plus that from material accreted later. As the putative accretion disk is very likely to have an angular momentum vector oriented in the same direction as that of the young star around which it orbits, we can use the instantaneous angular momentum vector of each sink particle to define the axis of the unresolved accretion disk.\\
\indent Once an axis has been defined, the simplest way to achieve a polar wind is to modulate the initially--isotropic distribution of wind momentum using a function proportional to the cosine of the polar angle, raised to some power $n$ which determines the degree of collimation. There is no real physics involved in this -- it is simply a convenient device by which we can explore the general importance of collimation. Modulating the wind momentum flux in this way would result in a lower radiated wind power.  We perform simulations in which we compare the effects of collimated and uncollimated winds. In the case of collimated winds, we increase the source momentum fluxes so that the actual radiated wind power of the collimated and uncollimated wind sources is the same. For an arbitrary power $n$, we achieve this by multiplying the wind momentum flux by a normalizing factor $F$ given by
\begin{eqnarray}
F=4\pi\left[\int_{0}^{2\pi}d\phi\int_{0}^{\pi}cos^{n}\theta sin\theta d\theta\right]^{-1}.
\end{eqnarray}
\indent In our simulations in which we employ anisotropic winds, we
use $n=5$, so that the polar angle angle at which the winds flux falls
to half its maximum value (which gives a measure of the wind opening
half--angle) is $\sim29^{\circ}$, corresponding to a wind weakly
collimated by a disk. By comparison with a simulation in which the
winds are not intrinsically collimated, we can study separately the
effects of small--scale collimation by circumstellar disks, and
collimation on larger scales by accretion flows or other structures in
the gas. We term this collimation on larger scales 'extrinsic'.\\
\indent The Monte Carlo technique is relatively computationally expensive, although the algorithm can be parallelised efficiently. The overhead from the mometum injection wind code in calculations involving a single wind source depends quite strongly on how many particles are in the working face. In the three tests detailed above, involving single sources and run on a single CPU machine, the fractional runtime consumed by the wind code ranged from negligible in the early stages of simulations to up to $\sim80\%$ towards the end. In the calculations detailed below in Section 5, run on $4$--CPU machines and in which $6-12$ sources are involved, the simulations with winds active took $\sim5$ times longer to run than those without. Nothwithstanding this considerable overhead, the momentum--injection code is very much faster than the particle--injection code and enables calculations to continue for timescales on the order of cluster freefall times so that the efects of winds on early cluster evolution, for example gas expulsion and cluster unbinding, may be studied.\\

\section{Particle winds and intrinsically--collimated outflows}

We apply our particle-injection method to test the intrinsic
collimation of isotropic winds.
In order to do this we make use of the massive stars formed in previous numerical simulations reported in  \cite{2002MNRAS.336..659B} and  
\cite{2003MNRAS.343..413B}.  In the former, we model only the central part of the
cluster containing 327 stars of total mass 151 \solmas\ embedded in 371 \solmas\ of gas.
The wind is from a 30 \solmas\ star located at the centre of the system and is surrounded by a remnant
filament of gas (see first panel of figure~\ref{outflowBB02}). The wind is emitted just beyond the sink-radius of 10 AU with a mass outflow rate of $1 \times 10^{-5}$ \solmas yr$^{-1}$, a wind velocity of 1000 km s$^{-1}$ and an internal sound speed of 10 km s$^{-1}$.
The evolution was followed for a total of 85 years, by which time the
wind consisted of 90,000 particles and there were 77,000 regular gas
particles. The total outflow distance is only 5600 AU, corresponding to an average expansion velocity of $\sim300$km s$^{-1}$. If taken at face value, this would imply that the wind particles are slowed by interaction with a quantity of surrounding gas equal to only approximately four times their own mass and that the interaction of the two fluids is not well resolved. We find in fact that the wind particles travel almost ballistically until they strike the sorrounding gas, rapidly depositing almost all their momentum and coming virtually to a halt. There is no interparticle penetration and the interaction of the wind with the surrounding gas is well resolved. It is the short timesteps required to integrate this process that makes it impractical to evolve these calculations for long timescales. We also find, however, that the wind particles find and fill holes in the surrounding gas. Since low density regions are not well resolved in SPH, this process may not always be treated accurately.\\

\begin{figure}
\centerline{\hbox{\includegraphics[width=0.225\textwidth]{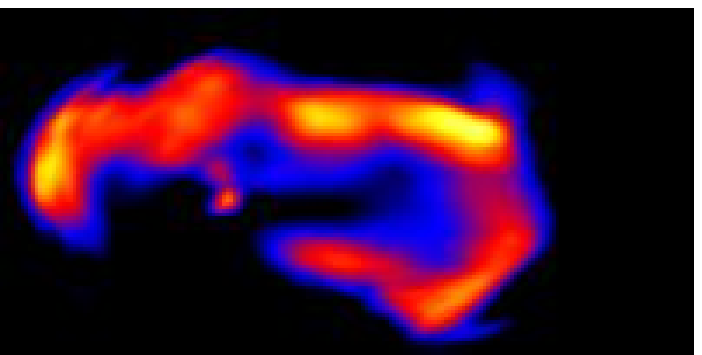} 
\includegraphics[width=0.225\textwidth]{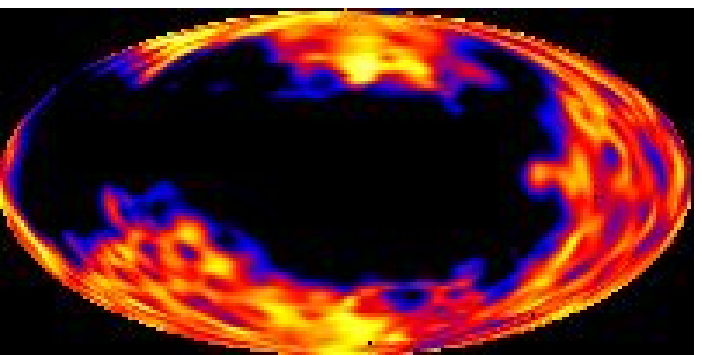}}}
\vspace{0.5mm}
\centerline{\hbox{\includegraphics[width=0.225\textwidth]{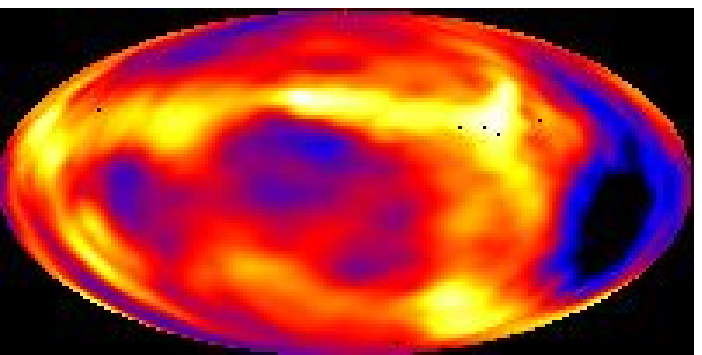} 
\includegraphics[width=0.225\textwidth]{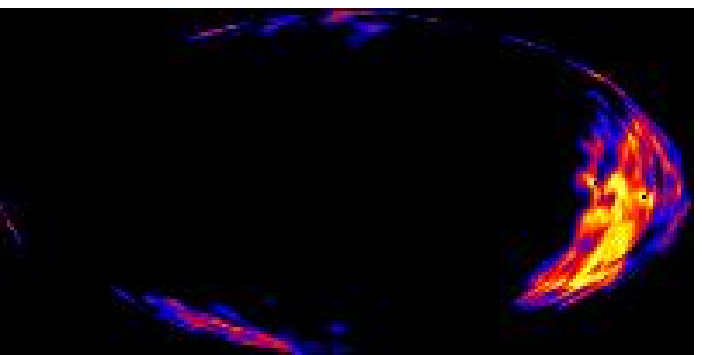}}}
\caption{Aitoff projections of the cold cluster  gas (left) and the hot wind (right) are shown
divided into regions close to the star ($<500$ AU, top) and further away ($> 500$ AU, bottom). 
We see that the cluster
gas impedes the wind in certain directions, thereby collimating it into the channels where
the column density of the cluster gas is smallest.}
\label{aitofwind}
\end{figure}

\begin{figure*}
\centerline{\psfig{figure=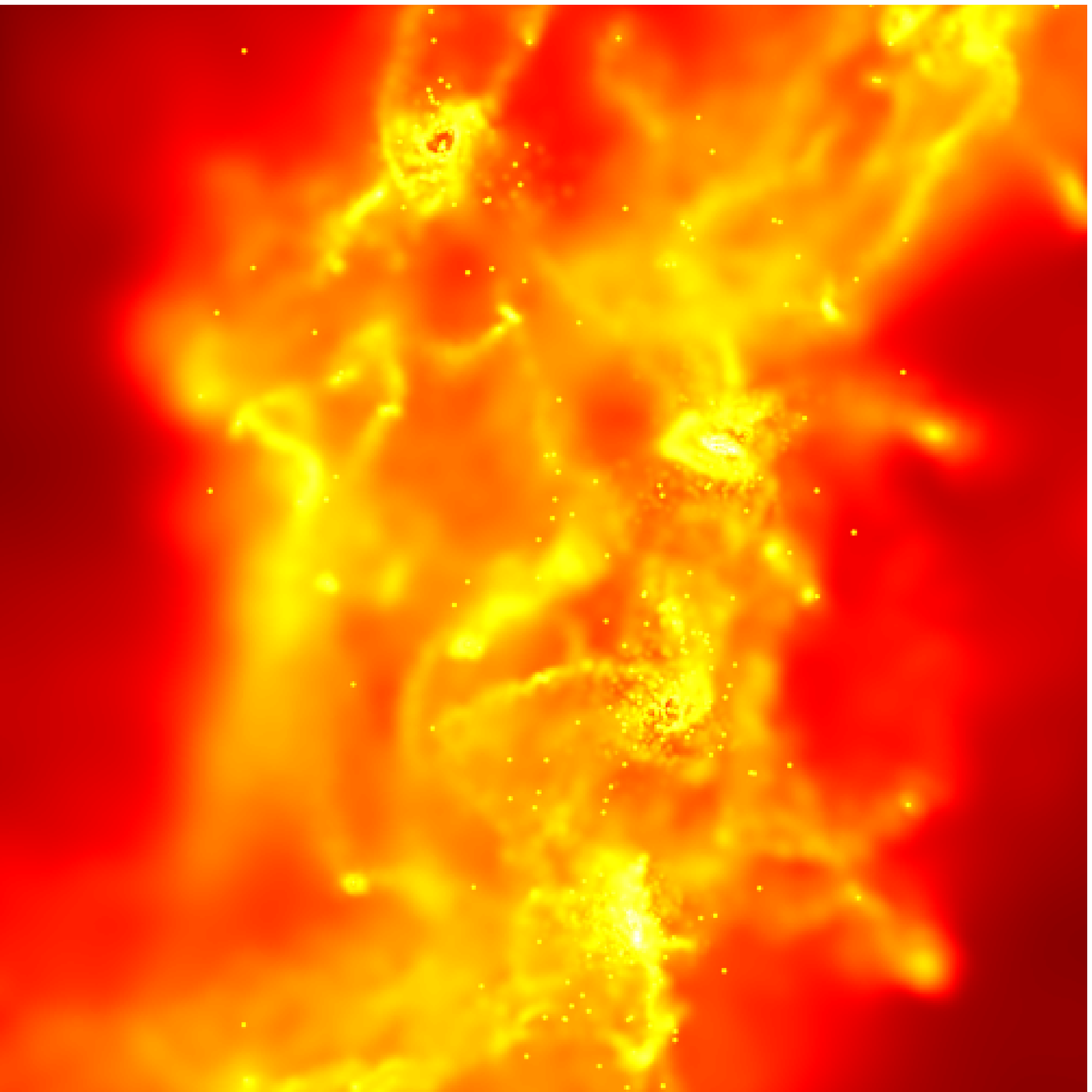,width=6.truecm,height=6.truecm}
\psfig{figure=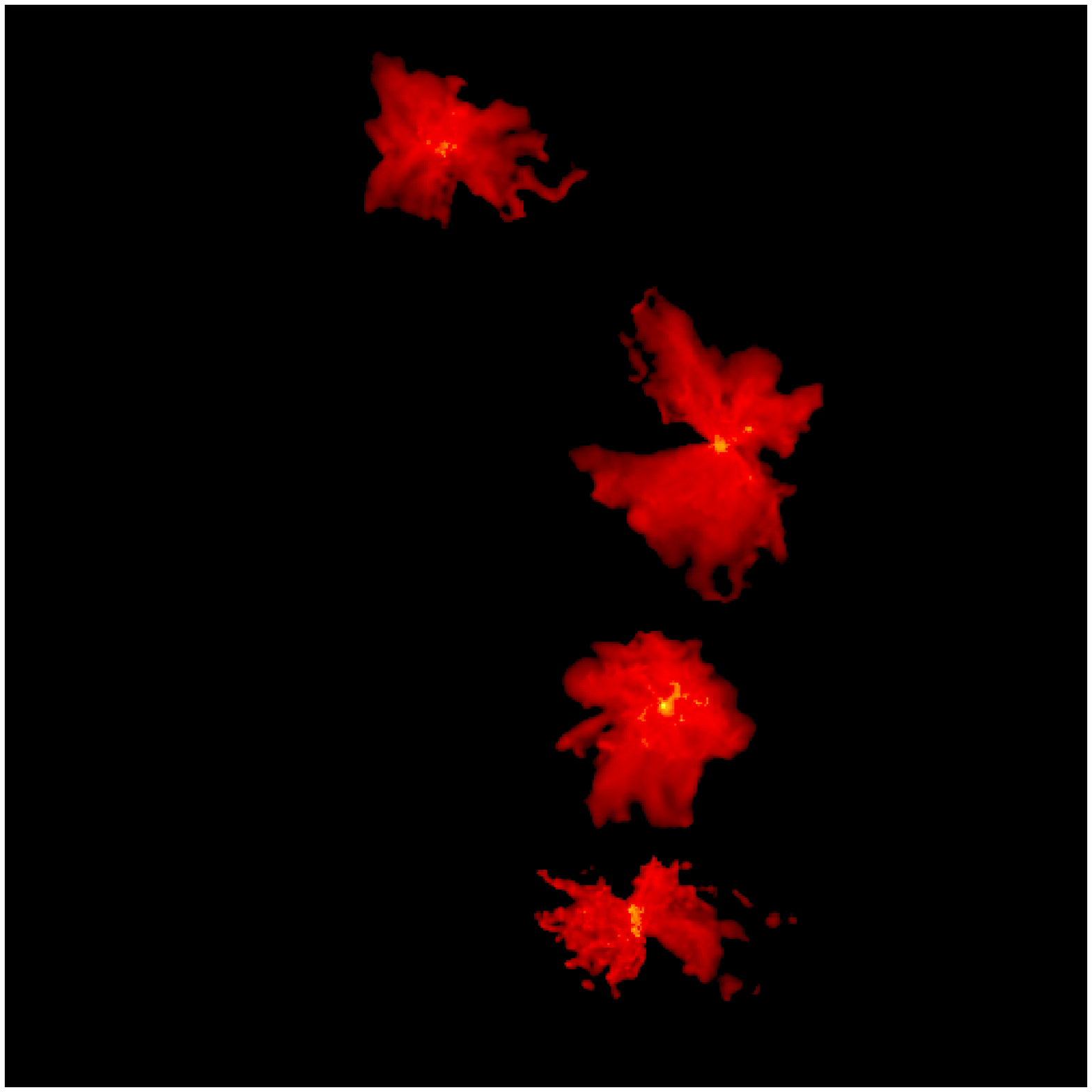,width=6.truecm,height=6.truecm}}
\caption{\label{outflow6BBV03} The winds from six massive stars (right) are shown
as they interact with the cluster environment (left). Initial conditions taken from the self-consistent cluster formation simulation of Bonnell et. al~(2003). Note that only four winds are visible as  two of the six sources are in the same binary system. The winds are shown 440 years after they were launched
and the box is 0.5 pc on a side.}
\end{figure*}

Figure~\ref{outflowBB02} shows the evolution of the  wind as it encounters the surrounding medium. The filamentary structure of the surrounding gas acts to impede the progress
of the wind, which eventually breaks out only in one direction producing a {\sl unipolar} outflow
that is highly collimated. This collimation of a spherical wind due simply to the external
medium occurs even in the absence of any rotationally supported disc
in the simulation, being caused instead by the filamentary gas structure in which the massive star is embedded.
Aitoff projections shown in Figure~\ref{aitofwind} show that there is
no particle interpenetration present in the simulation. Instead, the
wind is halted when it interacts with the cold cluster gas. It is only
in directions in which the column density of the surrounding gas is
low that the wind is able to clear a channel through which it can escape.

Incorporating such winds into larger scale simulations is also possible although at the expense of
no longer following the long-term evolution of the full system. We take the same initial conditions as 
detailed below (from \cite{2003MNRAS.343..413B}),  and inject particle-based winds. 
All stars whose mass $M_{*}$ exceeds $10$ M$_{\odot}$ are considered to be wind sources. Each such object is first assigned a wind mass loss rate $\dot{M}$ according to the expression
\begin{eqnarray}
\dot{M}=10^{-5}\left(\frac{M_{*}}{30\textrm{M}_{\odot}}\right)^{4}\textrm{ M}_{\odot}\textrm{ yr}^{-1}.
\label{eqn:mom_flux}
\end{eqnarray}
All sources are taken to have a wind terminal velocity $v_{\infty}$ of $10^{3}$ km s$^{-1}$. Wind particles are injected at the sink radii (200 AU) of the six stars whose mass exceeds $10$ M$_{\odot}$
and the evolution is followed for $\approx 440$ years (Fig.~\ref{outflow6BBV03}).  We note that even without any material
within 200 AU, the flows are likely to emerge as bipolar due to the
collimation by structures present at larger scales in the cloud.

\begin{figure}
\centerline{
\psfig{figure=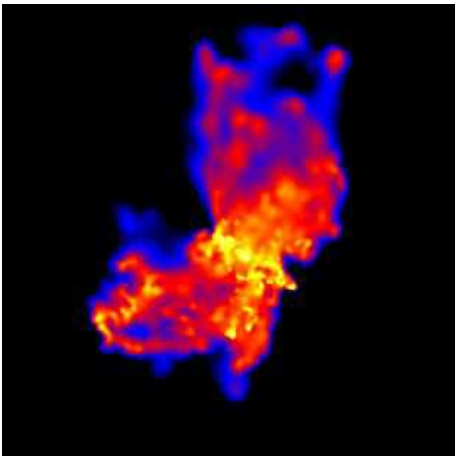,width=6.truecm,height=6.truecm}}
\caption{\label{windMS1} The winds from the most massive star from a higher resolution simulation where circumstellar structures are resolved down to 50 AU. The discs present at these scales are able to collimate the outflows into bipolar morphologies. 
The figure shows the hot gas present in the wind in a projection where the circumstellar disc is nearly edge-on. The image shows a region approximately 4000 AU on a side.}
\end{figure}

Figure~\ref{windMS1} shows a higher resolution simulation of the wind from the most massive star
in the above calculation. In this case, we resolve structures down to 50 AU from the star while the wind is launched at 20 AU. This allows the resolution of circumstellar discs in the simulation (see Clark et. al 2007) that further collimate the stellar winds. This results in better collimation of the outflows into classical bipolar morphologies. The bright ring of material in the wind shows the
interaction with the circumstellar disc that is providing the collimation of the wind.

The above results demonstrate that any spherical wind is likely to be collimated by the external
environment and often by structures on small scales which are
difficult to resolve in simulations. Unfortunately, using direct particle winds is computationally very expensive, so such simulations cannot be followed for large fractions of the lifetime of a typical protocluster. In addition, the filling of holes in the surrounding gas by the wind particles raises issues of resolution, since these regions are by definition not well resolved and may not be true holes. These issues are exacerbated by the use of two different fluids with different temperatures, particle masses and smoothing lengths. One improvement, which we do not incorporate here, is to set smoothing lengths based on the requirement that a particle's smoothing kernel must contain the canonical number of fifty neighbours of particles of the same fluid type (i.e. `wind' or `regular'). We do not attempt this extension here, since the more basic problem of the short timesteps required by the particle injection scheme drove us to develop the momentum-driven wind method in order to explore the larger-scale and longer-term effects of feedback from massive stars on the forming protocluster. 

\section{Winds and Protocluster evolution}
The timescales for which we can compute the evolution of the particle--driven winds are necessarily short. In order to follow the long-term evolution of a forming protocluster, we use our momentum-driven
winds method to follow the feedback from the most massive stars in a self--consistent simulation.

\subsection {Initial Conditions}
\begin{figure}
\includegraphics[width=0.5\textwidth]{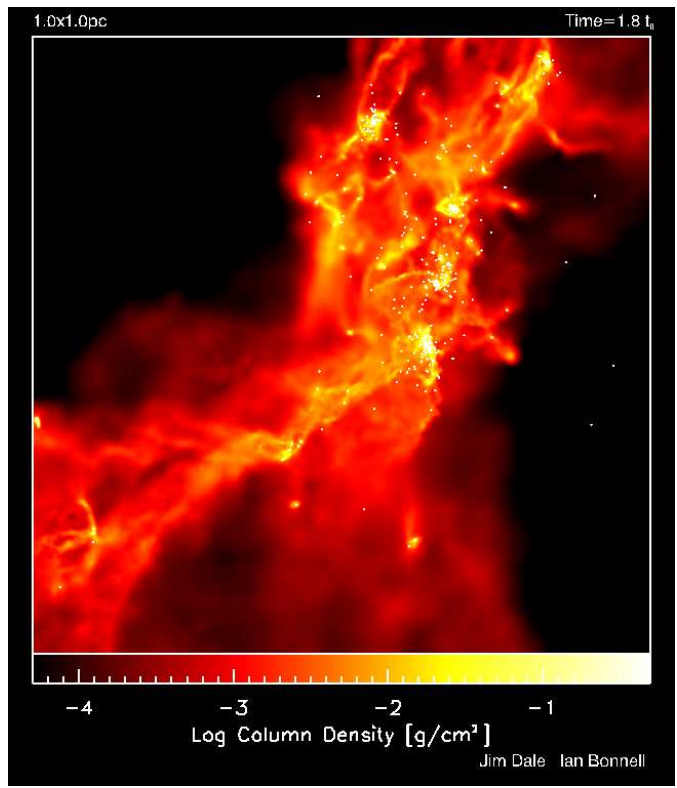}
\caption{Column--density map projected along the $z$--axis of the initial conditions for the momentum--injection wind simulations presented in this paper.}
\label{fig:wind_init}
\end{figure}

\begin{figure}
\includegraphics[width=0.5\textwidth]{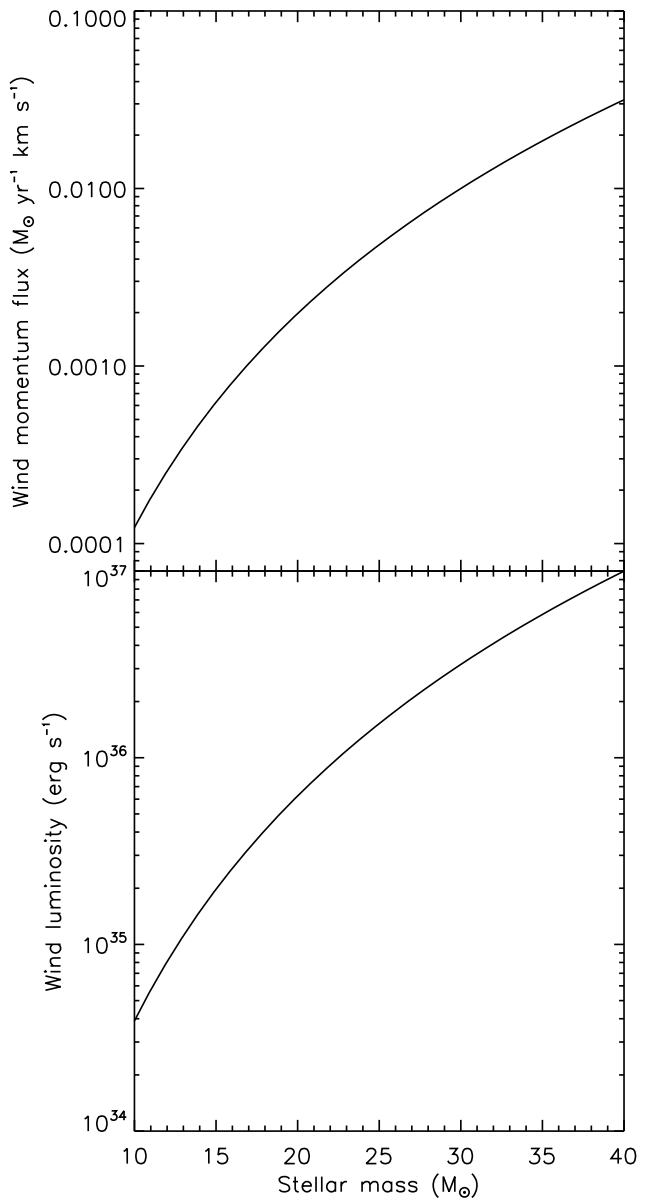}
\caption{Source wind momentum fluxes and wind luminosities as a function of mass as given by Equation \ref{eqn:mom_flux}.}
\label{fig:wind_sourcelum}
\end{figure}

We take as our initial conditions simulations performed by \cite{2003MNRAS.343..413B}. These authors studied the evolution of a $10^{3}$ M$_{\odot}$ molecular cloud, initially $1$ pc in diameter with a gas temperature of $10$ K. The cloud was initially spherical, but seeded with a divergenceless Gaussian supersonic turbulent velocity field with a power spectrum $P(k)\propto k^{-4}$, where $k$ is the wavenumber of the velocity perturbation. The total kinetic energy of the cloud of the cloud was set equal to the modulus of the gravitational potential energy, making the cloud marginally bound. The thermal Jeans mass of the cloud was $1$ M$_{\odot}$ and an isothermal equation of state was used throughout.\\
\indent \cite{2003MNRAS.343..413B} followed the evolution of the protocluster for $4.5\times10^{5}$ yr, equivalent to approximately $2.6$ freefall times. They observed that star formation in the system was hierarchical. They found that the gas initially fragmented at several locations, producing a number of small high--density clusters after $\approx2.4\times10^{5}$ yr ($1.4$ t$_{ff}$). Over the next freefall time or so, these smaller clusters fell in towards each other and merged, forming a single large, centrally--condensed cluster containing $\approx400$ stars.\\
\indent \cite{2003MNRAS.343..413B} point out, however, that several stars whose mass exceeds $10$ M$_{\odot}$ form by the end of the simulation, and their mechanical and radiative feedback may be expected to have a dramatic effect on the subsequent evolution of the system. In the calculations presented here, we model the mechanical feedback, i.e. the winds from these few massive stars.\\
\indent We turn on winds at $3.1\times10^{5}$ yr, $\approx1.8$
freefall times into the cluster's evolution. By this stage, the
hierarchical star formation process has produced five well--defined
subclusters and there are six stars whose mass exceeds $10$
M$_{\odot}$. The total number of stars is $377$. The total stellar
mass is $\approx411$ M$_{\odot}$, so that the star formation
efficiency at this stage of the system's evolution is
$\approx41\%$. We show a column--density map of the cluster at this
epoch in Figure \ref{fig:wind_init}. Stars are shown as white dots. We
perform three runs -- a control run which is allowed to evolve without
the action of winds and two `feedback' runs in which isotropic or
anisotropic winds are active.  We again treat any object whose mass exceeds $10$ M$_{\odot}$ as a wind source. All wind sources have mass--loss rates given by Equation \ref{eqn:mom_flux} and wind terminal velocities of $1000$ km s$^{-1}$. The momentum fluxes $\dot{M}v_{\infty}$ and wind luminosities $\dot{M}v_{\infty}^{2}/2$ of our sources as a function of mass are shown graphically in Figure \ref{fig:wind_sourcelum}.\\

\begin{figure}
\includegraphics[width=0.45\textwidth]{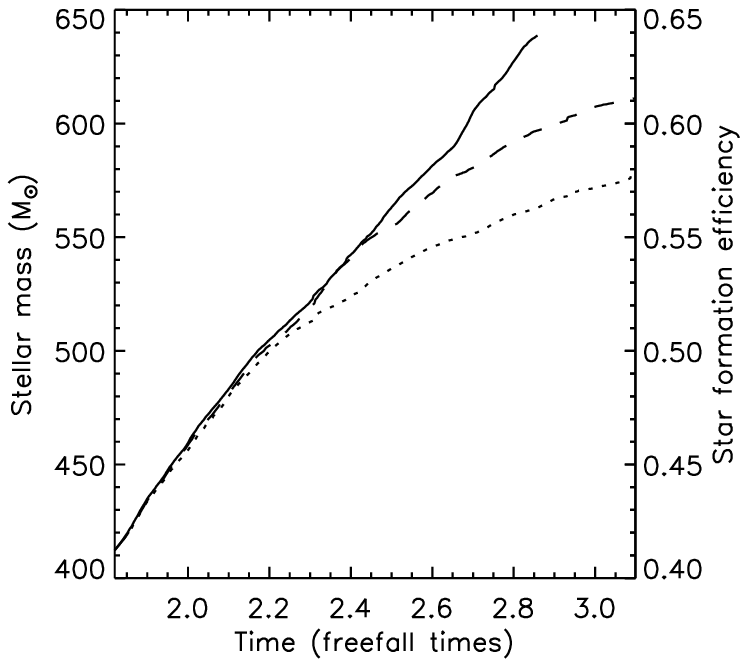}
\caption{Evolution with time of the total stellar mass and star--formation efficiency in the simulation without feedback (solid line), the simulation with isotropic wind feedback (dotted line) and the simulation with intrinsically--collimated wind feedback (dashed line).}
\label{fig:wind_sfe}
\end{figure}
\indent We follow the evolution of the feedback runs and the control run for a further $1.3$ freefall times until the cluster is $\sim3.1$ freefall times old. We followed the control run for a slightly shorter time, until an age of $\sim2.9$ freefall times.\\
\section{Results II: proto-cluster evolution}
\begin{figure*}
\includegraphics[width=\textwidth]{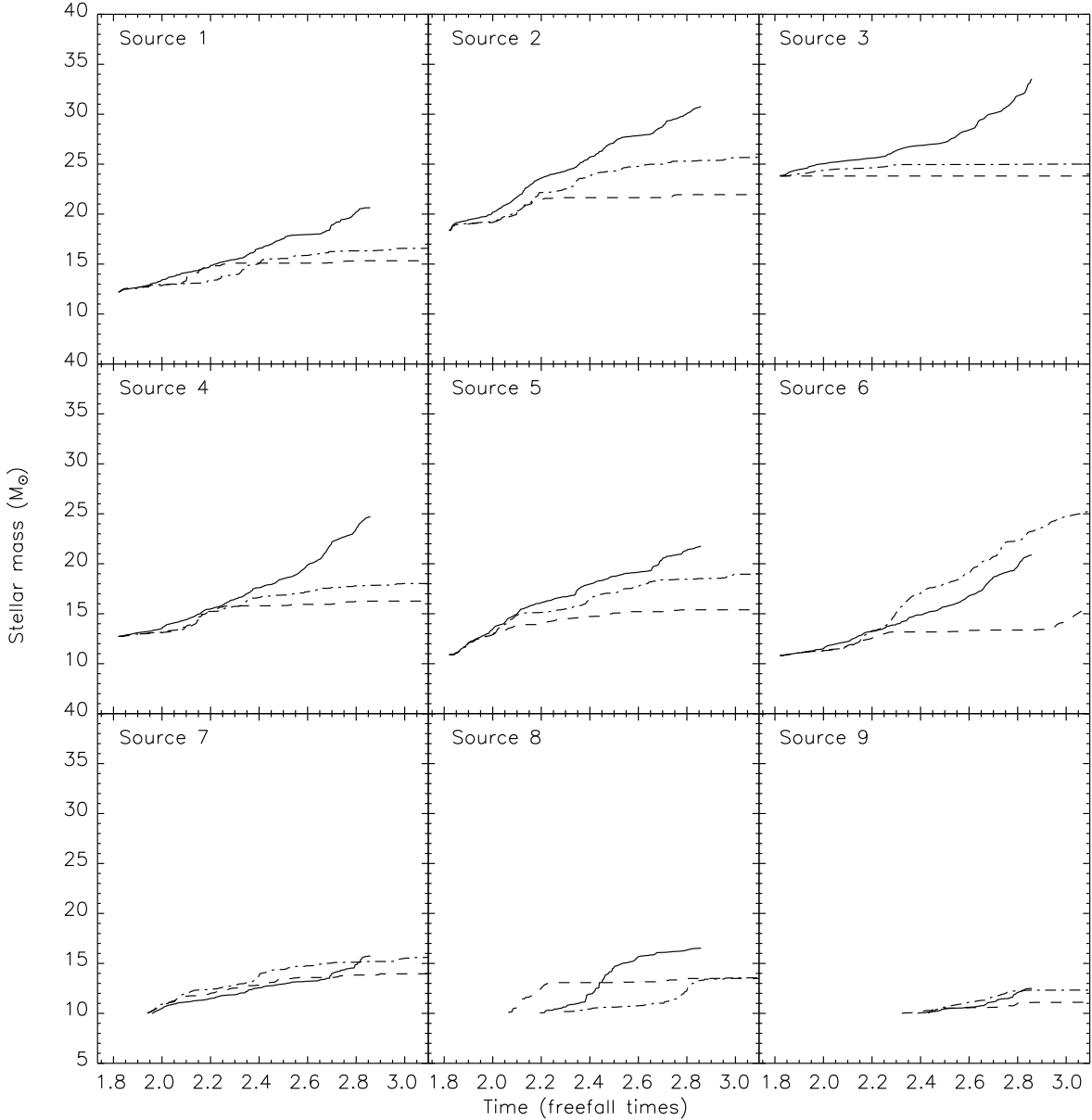}
\caption{Evolution with the time of the masses of the nine most massive objects in the control run (solid lines), compared with the evolution of the same objects in the isotropic (dashed line) and anisotropic (dot--dashed line) winds runs.}
\label{fig:wind_sources}
\end{figure*}

\begin{figure}
\includegraphics[width=0.5\textwidth]{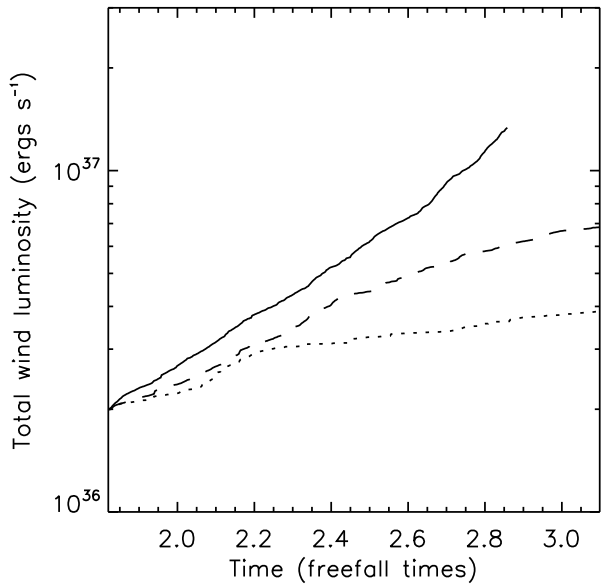}
\caption{Evolution with time of the total instantaneous wind luminosities in the control run (solid line, assuming all winds sources were suddenly turned on at each instant), in the simulation with isotropic wind feedback (dotted line) and the simulation with intrinsically--collimated wind feedback (dashed line).}
\label{fig:wind_lum}
\end{figure}

\begin{figure}
\includegraphics[width=0.5\textwidth]{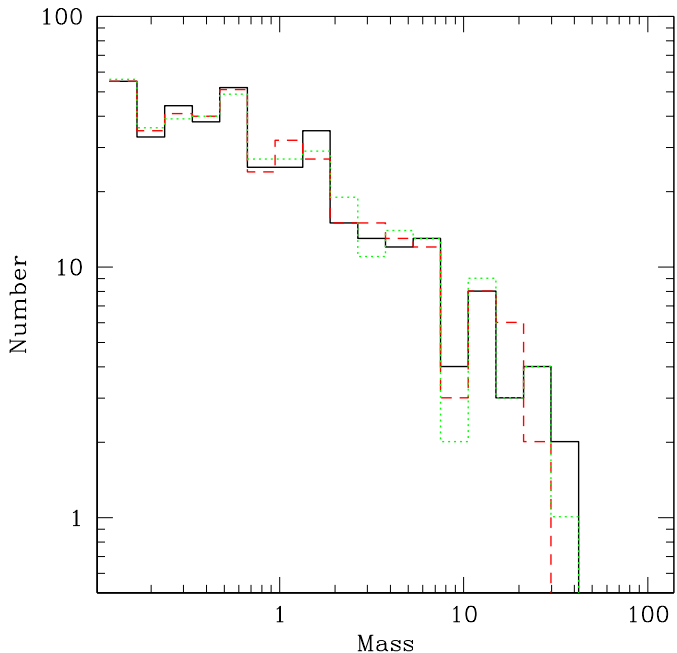}
\caption{The IMFs of the control run (solid black line), the isotropic winds run (solid red line) and the intrinsically--collimated winds run (dotted green line). Masses are given in solar masses.}
\label{fig:imf}
\end{figure}

\begin{figure*}
\includegraphics[width=\textwidth]{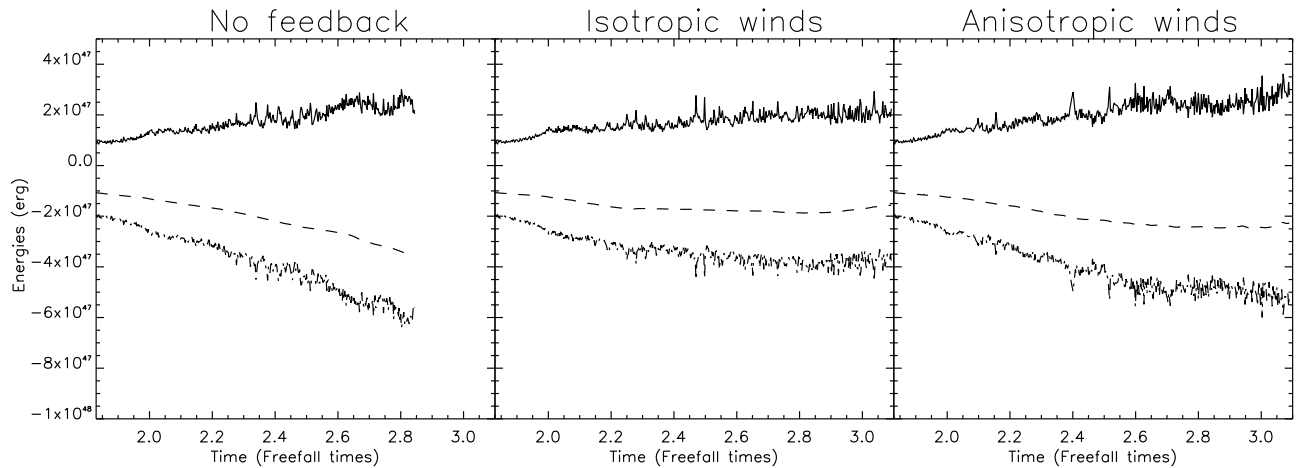}
\caption{Evolution with time of the global kinetic (solid line), gravitational potential (dot--dashed line) and total (dashed line) energies in the runs without feedback (left panel), with isotropic winds (centre panel) and with anisotropic winds (right panel).}
\label{fig:wind_nrg}
\end{figure*}

\begin{figure*}
\includegraphics[width=\textwidth]{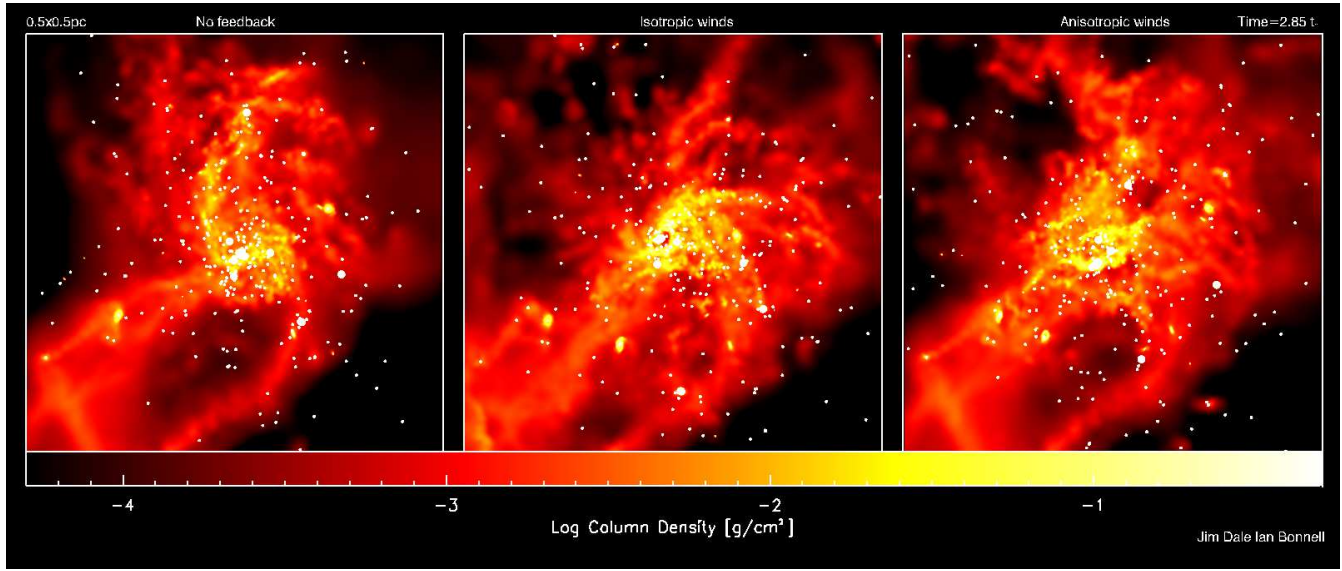}
\caption{Column density maps projected along the $z$--axis of the run without feedback (left panel), the run with isotropic winds (centre panel) and the run with anisotropic winds (right panel). White dots represent stars with the large dots denoting stars whose masses exceed $10$ M$_{\odot}$ (i.e. the stars which are wind sources, or those that would be if winds were included in the run without feedback).}
\label{fig:wind_screenshots}
\end{figure*}

\indent In Figure \ref{fig:wind_sfe} we plot the evolution with time of the total stellar mass and the star--formation efficiency (SFE) in our three simulations. We see that the SFE increases approximately linearly with time in the run without feedback, as would be expected for a gravitationally--bound isothermal system. By contrast, in both runs without feedback, although the SFE also increases linearly with time in the early parts of the simulations, the increase begins to tail off. The departure from the behaviour of the no--feedback calculation is gradual in both cases and occurs later in the run with anisotropic winds (at $\sim2.4$ t$_{ff}$) than it does in the run with spherically--symmetric winds (at $\sim2.2$ t$_{ff}$). The winds do not induce the formation of many new stars and do not produce any large cores which subsequently collapse monolithically to form massive stars. Very few new stars form in any calculation, so the discrepancy between the SFE of the three runs at any given time is almost entirely due to different accretion rates of pre--existing stars.\\
\indent This effect is clear when accretion on the the wind sources themselves is examined. In Figure \ref{fig:wind_sources}, we plot the accretion histories of the nine most massive stars in the run without feedback and compare their accretion histories with the same objects in the wind--affected calculations. In all cases, the effect of these stars' winds is to cut the rate at which they accrete. Source 3, whose initial mass is $\approx 23$ M$_{\odot}$, in particular accretes almost nothing in the isotropic winds run and very little in the anisotropic winds run during the $1.3$ t$_{ff}$ ($3.3\times10^{5}$) yr duration of our simulation. This is an excellent example of the self--regulation of feedback. The same effect is seen to varying degrees in all of Sources 1--5. Sources 6--9, however, are different in that their masses in the winds runs may be larger or smaller than their counterparts in the control run. It is also evident from Source 7 that accretion onto a given object may be strongly increased by the action of nearby winds, since this source crosses the $10$ M$_{\odot}$ threshold much earlier in the anisotropic winds run than in the control run. It is clear from Figure \ref{fig:wind_sources} that, although the effect of winds on the global accretion rate in the cluster may be simply described, the effect on the accretion rate of individual objects, whether they are wind sources or not, is much more complicated.\\
\indent In Figure \ref{fig:wind_lum}, we explore the self--regulation issue by plotting the wind luminosity from all stars with $M_{*}>10$ M$_{\odot}$ (as computed using Equation \ref{eqn:mom_flux}) as a function of time in the wind runs and comparing it to the total flux that all such sources in the windless run \textit{would} have if they were turned on at that instant. We see that the wind luminosity levels off in the isotropic wind run as accretion onto the wind sources slows almost to a halt. In contrast, the fictional wind luminosity in the run without feedback continues to rise almost linearly as accretion proceeds unchecked. The anisotropic winds run lies between these extremes, as accretion onto the winds sources is not slowed as efficiently.\\
\indent Since the accretion rates onto stars are clearly affected by winds, we examined the global effect of winds on the initial mass functions (IMFs) of our model clusters, depicted in Figure \ref{fig:imf}. The similarity of the mass functions below $\sim8$M$_{\odot}$ demonstrates that winds have little effect on the IMF of low-- and intermediate--mass stars in these simulations. The IMFs do differ somewhat above $\sim8$M$_{\odot}$, approximately the same range of masses as the objects considered to be wind sources ($>10$M$_{\odot}$). Winds, as also shown in Figure \ref{fig:wind_sources}, slow or stop the accretion of the wind sources themselves and can prevent them evolving to higher masses. This is seen in Figure \ref{fig:imf} where the two most massive sources in the isotropic winds run (the red dashed line) lie in the mass range $20-30$M$_{\odot}$, whereas these objects have evolved to be $30-40$M$_{\odot}$ in the run without winds (the solid black line). It also appears that a few objects in the isotropic winds run have piled up in the $10-20$M$_{\odot}$ mass bin, instead of evolving to higher masses as in the other runs. The IMF in the isotropic winds run thus appears slightly steeper at the high--mass end, but this interpretation should be treated with some caution, as it is based on $\sim10$ objects. The high--mass IMF of the run with intrinsically--collimated winds is intermediate between the two other runs. As mentioned earlier, winds do not induce the formation of may new stars and do not result in the formation of massive clumps giving rise to massive stars -- the massive stars in our calculations acquire almost all of their mass through accretion. We find therefore that the IMF in these calculations is still generated almost entirely by competitive accretion with little interference from winds.\\
\indent To explore the effect of the winds on the global dynamics of the cluster, we plot in Figure \ref{fig:wind_nrg} the evolution of the kinetic, gravitational potential and total energies of our cluster (the thermal energies in all runs are both constant and negligible in comparison to the other forms of energy). The short--timescale fluctuations in the kinetic and gravitational energies in all cases are due to close stellar interactions. In the run without feedback, we see that there is a general increase with time of global kinetic energy, but that this is more than counterbalanced by the global increase in magnitude of the gravitational energy, so that the system becomes more bound as time progresses. This behaviour is what one would expect from an initially--bound isothermal system. In the two runs in which winds are included, we see rather different behaviour. In both cases, the kinetic energy also increases with time. Towards the end of the run with anisotropic winds, the rate of increase becomes greater than in either of the other calculations. Additionally, in both simulations including winds, although the gravitational energy initially decreases in a similar manner to that in the run without feedback, the decrease is slowed somewhat. This is due to expulsion of material by the stellar winds, resulting in a corresponding drop in the gravitational binding energy. The overall result is that the total energy in the winds runs levels off and starts to become less negative at $\sim3.0 t_{ff}$. Both the isotropic and anisotropic winds are unbinding the model cluster.\\
\indent In Figure \ref{fig:wind_unbd}, we plot the evolution with time of the quantity of unbound mass (defined as the sum of the masses of all sink-- or SPH--particles with positive overall energy) in our three calculations.\\
\indent We see that the quantity of unbound mass is small ($<10\%$) and roughly constant in the run without feedback, as might be expected. The run in which isotropic winds are acting (the dotted line in Figure \ref{fig:wind_unbd}) shows an unbound mass fraction increasing approximately linearly from the beginning of the simulation. The increase is not great however, with only $\sim15\%$ of the cluster being unbound after winds have been acting for over a freefall time. Once again, the run including anisotropic winds shows behaviour intermediate between the other two calculations. For a period of $\sim0.6$ freefall times after the ignition of the anisotropic wind sources, they have almost no effect on the quantity of unbound mass in the cluster, as compared with the run in which winds of any kind are absent. However, after $\sim0.6$ freefall times, the quantity of unbound mass in the anisotropic winds run also begins to rise linearly at approximately the same rate as in the run with isotropic winds, to a similarly modest value of $\sim12\%$ after $\sim1.3$ freefall times.\\
\indent A comparison of the morphologies of the densest regions of the three model clusters is shown in Figure \ref{fig:wind_screenshots}. Surprisingly, the morphologies of the simulations are rather similar, so much so that they are difficult to tell apart. There is no well--defined wind bubble in either of the two runs in which winds are included. This appears to be due to the fact that the wind sources are moving around at high velocities into and then out of dense gas, rather than sitting in one place. They therefore do not have the opportunity to build up a single bubble. Accretion into the central regions of the cluster also contributes to the inability of the massive stars to construct a wind bubble. Instead of a largely radial outflow emanating from the most massive stars, gas is being driven out of the cluster anisotropically. The only notable difference between the run without feedback and the runs with winds are that the winds runs have quantities of gas at higher densities than exist in the control run. This is due to shock compression of some of the gas by the stellar winds.\\
\section{Discussion}
\begin{figure}
\includegraphics[width=0.45\textwidth]{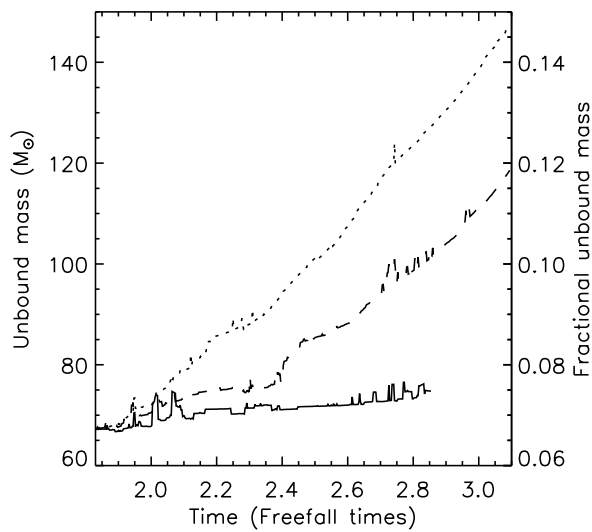}
\caption{Evolution with time of the quantity of unbound mass in the control run (solid lines), compared with the evolution of the same objects in the isotropic (dashed line) and anisotropic (dot--dashed line) winds runs.}
\label{fig:wind_unbd}
\end{figure}

\begin{figure*}
\includegraphics[width=\textwidth]{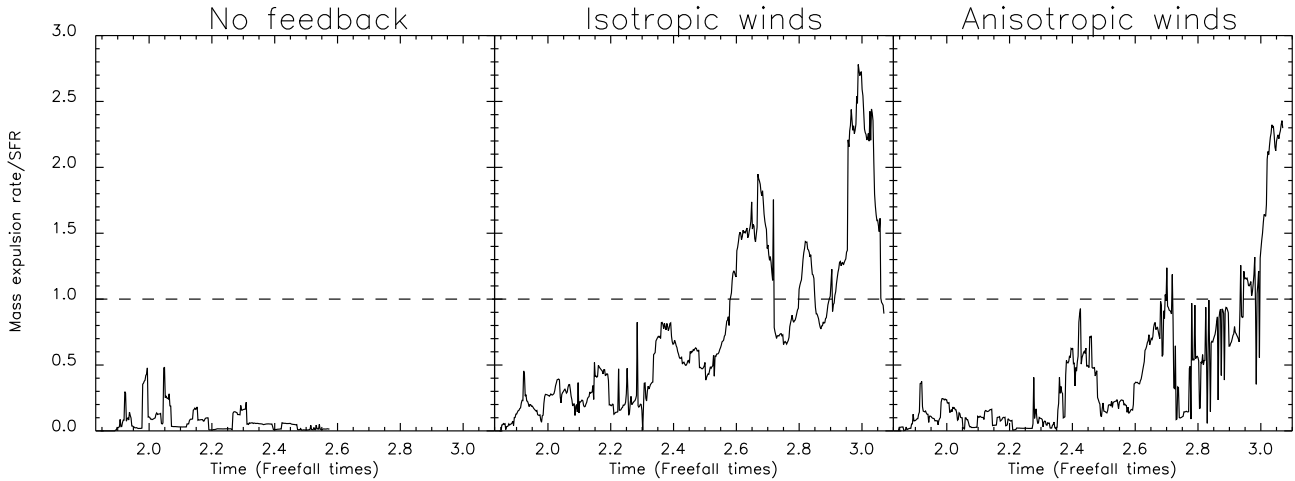}
\caption{Evolution with time of the ratio of the mass expulsion rate to the star formation rate (both measured in units of solar masses per initial global freefall time) in the control run (left panel), in the isotropic winds run (centre panel), and in the anisotropic winds run(left panel).}
\label{fig:wind_gen}
\end{figure*}

\indent If one thinks of accretion and expulsion from the system as two sinks of the remaining gas in our model cluster, one can see which of the two is dominant from the ratio of the rate at which material is becoming unbound from the system to the star formation rate (SFR). In Figure \ref{fig:wind_gen} we plot this ratio as a function of time for our calculations. If the ratio is less than one (below the dashed line in the plots), gas is being converted to stars faster than it is being expelled, in which case feedback is not the dominant factor controlling the dynamics of the gas. Conversely, if this ratio exceeds one, gas is being expelled from the cluster faster than it is being converted to stars and feedback is dominant. Not surprisingly, in the run without feedback, the ratio is always less than one. In the calculations in which we included winds, the ratio is less than unity for most of the duration of our simulations, although shows a general upward trend as the simulations progress (note, however, that neither the global accretion rate nor the rate at which material becomes unbound are steady functions, so there are considerable deviations from this trend). In both the simulations including winds, after winds have been acting for $\sim$ a freefall time, the rate at which material is being expelled from the cluster generally exceeds the rate at which it is being accreted by existing stars or forming new ones, so feedback is in control of the gas dynamics. However, from Figures \ref{fig:wind_sfe} and \ref{fig:wind_unbd}, we know that this trend is not due to an acceleration in the rate at which the system is losing gas, but rather to a dropoff in the global gas accretion/star formation rate. $\sim1$ freefall time after winds are activated (and $\sim3$ freefall times into the lifetime of the protocluster), in both winds runs, gas is being unbound at an approximately constant rate of $\sim50$ M$_{\odot}$ per freefall time. Given that, in both cases, there is still $\sim400$ M$_{\sun}$ of gas remaining in the cluster in both simulations and the accretion/star formation rate is tailing off, we conclude that feedback is likely to take $\sim8$ freefall times to clear the system of gas by the action of winds alone.\\
\section{Conclusions}
\indent It is clear from these simulations that spherical stellar
winds can be collimated by structures near the wind sources at a wide
range of scales. Whether the winds are isotropic or intrinsically
anisotropic (i.e. collimated by some structure very close to the
source) has some influence on the effectiveness of the winds, but
largely to delay the point at which winds become important in
controlling the cluster dynamics and the star--formation process. We
found that our intrinsically--collimated winds had similar effects to
the isotropic winds, but delayed by $\sim0.2$ freefall times, a
relatively small fraction of the duration of our simulations. In
either case, winds can have a strong effect on the evolution of protoclusters.\\
\indent We have shown that winds can dramatically slow the process of conversion of molecular gas into stars. In our simulations, this was achieved largely by slowing or stopping accretion on existing stars, rather than preventing the formation of new stars, since few new stars formed in any of our simulations after the point at which wind sources were introduced ($\sim1.8$ freefall times into the evolution of the system). In particular accretion onto the more massive wind sources themselves, which constitute a significant fraction ($\sim20 \%$) of the total stellar mass, was severely curtailed by isotropic or anisotropic winds. Since the accretion rates onto the intermediate-- and low--mass stars were not so strongly affected, we found that winds had little discernible effect on the cluster IMF except at the high--mass end, at and above a mass similar to the threshold mass we chose to determine whether a star was considered to be a wind source or not. It is possible that the (slight) changes in the IMF are therefore influenced by our choice of this parameter, so we do not attempt to draw strong conclusions on this subject, except that winds appear to have little influence on the generation of the IMF by competitive accretion. Although there do not appear to be any instances of star formation induced by winds in our calculations, this may only be a consequence of the fact that the clusters are only marginally bound and that the wind sources are turned on relatively late, after much of the bound gas has already fragmented. It s possible that, in clusters that are more strongly bound, winds acting earlier in the cluster's evolution may encourage fragmentation and induce star formation.\\
\indent Additionally, the stellar winds from a few stars can come to dominate the dynamics of a whole cluster, so that the rate at which mass is being expelled from the cluster comes to exceed the rate at which mass is being converted to stars. In this case, the winds are beginning to unbind the cluster and an examination of the evolution of the kinetic, gravitational and total energies of our model clusters revealed that the action of winds is to stop and even reverse the decline in the cluster's total energy. However, the rates at which winds were able to expel gas from the clusters in our simulations were relatively low ($\sim50$ M$_{\odot}$ per freefall time, or $\sim5$ percent of the cluster's total mass per freefall time). This amounts to $\approx 3\times10^{-4}$M$_{\odot}$yr$^{-1}$. As can be seen by comparison with Figure \ref{fig:wind_sourcelum}, this is much less than the momentum fluxes of the sources, if the gas is expelled at velocities comparable to the cluster escape velocity of a few kilometres per second. We identify three reasons for this discrepancy. Firstly, gas is generally expelled from regions near the massaive stars, where escape velocities are $\sim100$km s$^{-1}$. Secondly, some of the gas leaving the cluster is moving at velocities in excess of the escape velocity. Thirdly, due to the inhomogenous gas distribution, some of the momentum flux from the sources is simply lost through holes in the gas. This result implies that the momentum flux of embedded wind sources alone may not be a good criteroin for determining whether a given protocluster can be dispersed by winds.\\
\indent The slow rate of gas expulsion led to an interesting situation in which our wind--influenced clusters still possessed large quantities of gas which it would take the winds nearly ten freefall times to expel, but were converting gas to stars very slowly. The time for which the clusters were forming stars at a the rate dictated by purely isothermal evolution (as observed in the control run without winds) would then be only $\sim20$ percent of the time for which they possessed molecular gas. If generally true, this would have important implications for calculations of star formation rates based on dividing the stellar mass of a system by an age derived from its oldest stars. If the cluster in question had spent a large fraction of its time forming stars very slowly because of the feedback effects observed in our winds runs, the inferred star formation rate would not be representative of the earliest stage of the cluster's evolution. More simulations are obviously required to determine the likelihood of this being true.\\
\indent We note that the morphology of the gas in our wind--influenced simulations prevents the formation of more--or--less spherical wind bubbles. In our simulations, the high relative velocities of the wind sources to each other and to the gas, and the continuing large scale accretion flows present in the model clusters prevented the formation of such structures. It is actually very difficult to tell which of our simulations include winds and which do not from the gas morphology alone, except that the calculations including winds exhibit larger quantities of dense shocked gas. It may be true therefore that it is not always easy to see the effects of winds on young (i.e $<1$Myr old) clusters.\\
\indent In this work, we have not included the effects of photoionising radiation. The wind sources in our calculations are massive enough to have strong ionising fluxes and these would also influence the evolution of the protocluster. It is not clear whether winds and photoionisation acting together help or hinder each other in unbinding stellar systems -- winds may help clear out dense material from around the massive stars, allowing ionising radiation to escape, but swept up shells of wind material may confine the HII regions at relatively small radii and prevent them having much effect. We leave this interesting question to a later paper.\\
\section{Acknowledgments}
JED acknowledges support from the University of Leicester's PPARC rolling grant and from the Wenner--Gren foundation. Some of the calculations reported in this paper were performed on the UK Astrophysical Fluids Facility machine at the University of Leicester.

\bibliography{myrefs}

\begin{thebibliography}{}

\bibitem[\protect\citeauthoryear{{Bate}, {Bonnell} \& {Price}}{{Bate}
  et~al.}{1995}]{1995MNRAS.277..362B}
{Bate} M.~R.,  {Bonnell} I.~A.,    {Price} N.~M.,  1995, \mnras, 277, 362

\bibitem[\protect\citeauthoryear{{Bonnell} \& {Bate}}{{Bonnell} \&
  {Bate}}{2002}]{2002MNRAS.336..659B}
{Bonnell} I.~A.,  {Bate} M.~R.,  2002, \mnras, 336, 659

\bibitem[\protect\citeauthoryear{{Bonnell}, {Bate} \& {Vine}}{{Bonnell}
  et~al.}{2003}]{2003MNRAS.343..413B}
{Bonnell} I.~A.,  {Bate} M.~R.,    {Vine} S.~G.,  2003, \mnras, 343, 413

\bibitem[\protect\citeauthoryear{{Capriotti} \& {Kozminski}}{{Capriotti} \&
  {Kozminski}}{2001}]{2001PASP..113..677C}
{Capriotti} E.~R.,  {Kozminski} J.~F.,  2001, \pasp, 113, 677

\bibitem[\protect\citeauthoryear{{Clark} \& {Bonnell}}{{Clark} \&
  {Bonnell}}{2004}]{2004MNRAS.347L..36C}
{Clark} P.~C.,  {Bonnell} I.~A.,  2004, \mnras, 347, L36

\bibitem[\protect\citeauthoryear{{Clark} \& {Bonnell}}{{Clark} \&
  {Bonnell}}{2005}]{2005MNRAS.361....2C}
{Clark} P.~C.,  {Bonnell} I.~A.,  2005, \mnras, 361, 2

\bibitem[\protect\citeauthoryear{{Dale}, {Bonnell}, {Clarke} \& {Bate}}{{Dale}
  et~al.}{2005}]{2005MNRAS.358..291D}
{Dale} J.~E.,  {Bonnell} I.~A.,  {Clarke} C.~J.,    {Bate} M.~R.,  2005,
  \mnras, 358, 291

\bibitem[\protect\citeauthoryear{{Elmegreen}}{{Elmegreen}}{2000}]{2000ApJ...53%
0..277E}
{Elmegreen} B.~G.,  2000, \apj, 530, 277

\bibitem[\protect\citeauthoryear{{Fukui} \& {Mizuno}}{{Fukui} \&
  {Mizuno}}{1991}]{1991IAUS..147..275F}
{Fukui} Y.,  {Mizuno} A.,  1991, in {Falgarone} E.,  {Boulanger} F.,   {Duvert}
  G.,  eds, IAU Symp. 147: Fragmentation of Molecular Clouds and Star Formation
  {A Comparative Study of Star Formation Efficiencies in Nearby Molecular Cloud
  Complexes}.
pp 275--+

\bibitem[\protect\citeauthoryear{{Garcia-Segura}, {Langer} \& {Mac
  Low}}{{Garcia-Segura} et~al.}{1996}]{1996A&A...316..133G}
{Garcia-Segura} G.,  {Langer} N.,    {Mac Low} M.-M.,  1996, \aap, 316, 133

\bibitem[\protect\citeauthoryear{{Hartmann}, {Ballesteros-Paredes} \&
  {Bergin}}{{Hartmann} et~al.}{2001}]{2001ApJ...562..852H}
{Hartmann} L.,  {Ballesteros-Paredes} J.,    {Bergin} E.~A.,  2001, \apj, 562,
  852

\bibitem[\protect\citeauthoryear{{Krumholz} \& {Tan}}{{Krumholz} \&
  {Tan}}{2007}]{2007ApJ...654..304K}
{Krumholz} M.~R.,  {Tan} J.~C.,  2007, \apj, 654, 304

\bibitem[\protect\citeauthoryear{{Lada} \& {Lada}}{{Lada} \&
  {Lada}}{2003}]{2003ARA&A..41...57L}
{Lada} C.~J.,  {Lada} E.~A.,  2003, \araa, 41, 57

\bibitem[\protect\citeauthoryear{{Lamers} \& {Cassinelli}}{{Lamers} \&
  {Cassinelli}}{1999}]{1999isw..book.....L}
{Lamers} H.~J.~G.~L.~M.,  {Cassinelli} J.~P.,  1999, {Introduction to stellar
  winds}.
Introduction to stellar winds / Henny J.G.L.M.~Lamers and Joseph
  P.~Cassinelli.~Cambridge ; New York : Cambridge University Press, 1999.~ ISBN
  0521593980

\bibitem[\protect\citeauthoryear{{Ostriker} \& {McKee}}{{Ostriker} \&
  {McKee}}{1988}]{1988RvMP...60....1O}
{Ostriker} J.~P.,  {McKee} C.~F.,  1988, Reviews of Modern Physics, 60, 1

\bibitem[\protect\citeauthoryear{{Steigman}, {Strittmatter} \&
  {Williams}}{{Steigman} et~al.}{1975}]{1975ApJ...198..575S}
{Steigman} G.,  {Strittmatter} P.~A.,    {Williams} R.~E.,  1975, \apj, 198,
  575

\bibitem[\protect\citeauthoryear{{Vishniac}}{{Vishniac}}{1983}]{1983ApJ...274.%
.152V}
{Vishniac} E.~T.,  1983, \apj, 274, 152

\bibitem[\protect\citeauthoryear{{Weaver}, {McCray}, {Castor}, {Shapiro} \&
  {Moore}}{{Weaver} et~al.}{1977}]{1977ApJ...218..377W}
{Weaver} R.,  {McCray} R.,  {Castor} J.,  {Shapiro} P.,    {Moore} R.,  1977,
  \apj, 218, 377

\end{thebibliography}

\label{lastpage}

\end{document}